# Bits About the Channel: Multi-round Protocols for Two-way Fading Channels

Vaneet Aggarwal[†] and Ashutosh Sabharwal[◇]

**Abstract**

Most communication systems use some form of feedback, often related to channel state information. In this paper, we study diversity multiplexing tradeoff for both FDD and TDD systems, when both receiver and transmitter knowledge about the channel is noisy and potentially mismatched. For FDD systems, we first extend the achievable tradeoff region for 1.5 rounds of message passing to get higher diversity compared to the best known scheme, in the regime of higher multiplexing gains. We then break the mold of all current channel state based protocols by using multiple rounds of conferencing to extract more bits about the actual channel. This iterative refinement of the channel increases the diversity order with every round of communication. The protocols are on-demand in nature, using high powers for training and feedback only when the channel is in poor states. The key result is that the diversity multiplexing tradeoff with perfect training and $K$ levels of perfect feedback can be achieved, even when there are errors in training the receiver and errors in the feedback link, with a multi-round protocol which has $K$ rounds of training and $K − 1$ rounds of binary feedback. The above result can be viewed as a generalization of Zheng and Tse, and Aggarwal and Sabharwal, where the result was shown to hold for $K = 1$ and $K = 2$ respectively. For TDD systems, we also develop new achievable strategies with multiple rounds of communication between the transmitter and the receiver, which use the reciprocity of the forward and the feedback channel. The multi-round TDD protocol achieves a diversity-multiplexing tradeoff which uniformly dominates its FDD counterparts, where no channel reciprocity is available.

[†]Department of Electrical Engineering, Princeton University, Princeton, NJ.
[◇]Department of Electrical and Computer Engineering, Rice University, Houston, TX.

# I. INTRODUCTION

Consider the feedback system shown in Figure 1, where the receiver first measures the channel and sends a feedback signal to the transmitter about the channel. The transmitter uses the channel information about the channel conditions to adapt its transmission strategy. A common feedback model involves quantizing the channel knowledge at the receiver into a finite number of bits. The finite level adaptation has been extensively studied for different adaptive transmission schemes like power control [1–18], rate control [8], beamforming [19, 20] and codebook adaptation [21]. A bulk of the work to-date assumes that either the channel knowledge at the receiver is perfect and/or the feedback channel is noise-free. The above body of work thus serves as an upper bound to the performance of a system, where channel information at the transmitter and receiver is approximate and potentially mismatched.

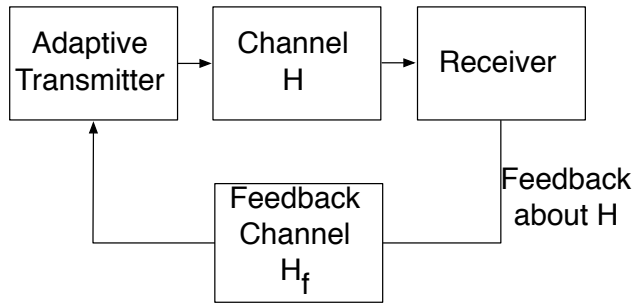

Fig. 1. A feedback based adaptive transmission system.

Imperfect channel knowledge at the transmitter and receiver is unavoidable in wireless communications since the channel $H$ (see Figure 1) is time-varying and hence has to be measured periodically by the receiver to assist in transmitter adaptation. To measure the channel, the transmitter sends a periodic finite length training signal, which is used at the receiver to form an estimate of the channel. The channel estimate is then quantized by the receiver and sent over a feedback channel which often is also a wireless link, and is thus prone to errors. In this scheme, a fundamental question is how much information *about* the channel can be extracted, which *both the transmitter and receiver can agree upon*. In short, the transmitter and receiver have to consent about the state of the channel by conversing over a noisy fading channel. In this paper, we make progress towards the above question for the case of power control, with the objective to minimize the outage probability.

If the channel information is known perfectly at the transmitter and the receiver, then the transmitter can perform power control to invert the effect of multiplicative fading channel. The inversion conserves power in good channel states by using less transmit power, while using more transmit power in poor channel conditions. If the transmitter knowledge is limited to a finite-bit approximation of the perfect receiver knowledge, then the quantized power control [2, 8, 12] is an approximation of the ideal power inversion. Much like the ideal channel inversion, the quantized power control delivers an average received signal-to-noise ratio of SNR by exploiting the fact that poor channel states are very rare and hence large peak powers are also used rarely.

For FDD systems, consider that the receiver quantizes the channel state and sends a feedback index consisting of $K$ levels. It was shown in [8, 12] that the diversity increases exponentially in $K$ for $mn > 1$ and linearly in $K$ for $mn = 1$. However, in [6], we show that if the receiver obtains the channel state information by training and the feedback from the receiver is sent over a noisy channel, feedback bits help in resolving the channel state till the noise floor becomes dominant. If the channel estimate is below the noise floor, the channel estimate does not have sufficient resolution to extract additional bits of information about the channel. This restricts the increase in diversity with feedback bits. We further showed that diversity equivalent of 1-bit perfect feedback and perfect channel state at the receiver can be achieved with 1 bit of noisy feedback with two rounds of training at the receiver. In this paper, we extend FDD results in two directions. First, we extend the achievability for $K$ bits of feedback showing gains for $r > \min(m, n) - 1$. In this regime, more bits of feedback help increase the diversity because more than one bit about the channel can be resolved if it is above noise floor, which can then be used for finer power control at the transmitter for higher multiplexing gains. Thus, more than one bit about the channel can be resolved using the first training.

Second, we extend the results in [6] to a quantized iterative protocol where we use $K$ rounds of training with $K - 1$ rounds of feedback to achieve the same diversity as $K$ levels of perfect feedback when the receiver knows perfect channel state information. The protocol uses $K$ rounds of training and $K - 1$ rounds of feedback to extract $\log_2(K)$ bits about the channel. The main idea in the proposed multi-round protocol is that the transmitter opportunistically sends higher training power when the previous round of training indicated that the channel is below the noise floor. The power used by the nodes to send a training/feedback/data symbol in a channel event is governed by the probability of the channel event itself, so that weighted average of the transmit power at the nodes meet the average power constraint. The above *on-demand use* of powers over many rounds helps increase the diversity beyond single shot estimation methods considered in all prior work. This is because the channel estimates at the nodes can be refined with increasing rounds of communication leading to larger diversities.

For TDD systems, the reciprocity in the forward and the backward channel can be utilized to achieve the gains in diversity. Due to reciprocity, if any of the nodes have perfect channel state information, the diversity gain is unbounded. However, if none of the nodes know channel state, reciprocity allows the receiver to detect if the transmitter has made an error in understanding the previous feedback signal. Thus, the receiver can correct transmitter's actions more rapidly compared to the case of FDD protocols, where such immediate error detection at receiver is not possible. The proposed protocol is able to achieve better diversity than the FDD results with 1.5 rounds. More precisely, this scheme achieves a diversity of $mn(mn+1) - (m+n-1)r$ for multiplexing $0 < r < \min(m,n)$. For a SIMO/MISO system, this strategy achieves the same diversity-multiplexing tradeoff as in the FDD system as the number of feedback levels go to $\infty$. The knowledge at the nodes can be further refined by more rounds of training and feedback. In general, a $(K-1).5$ round strategy is able to achieve a diversity of $mn(1 + mn + \cdots + (mn)^{K-1}) - (mn)^{K-2}(m+n-1)r$ for multiplexing $0 < r < \min(m,n)$. The FDD and TDD iterative schemes have the same diversity in the limit as $r \to 0$, but the TDD scheme leads to higher gains at all non-zero multiplexing gains.

The use of multiple rounds is similar in spirit to the improvement in error exponents by using the Schalkwijk-Kailath-like feedback-based coding scheme [22–24] for feedback in Gaussian channels. The multi-round strategies involve zooming into the region where there is some uncertainty. In the Schalkwijk-Kailath coding scheme, the transmitter starts with a coarse version of the message and then refines it with more rounds of feedback. In our scheme, the channel is unknown and the feedback is used to resolve the channel at the transmitter and the receiver. We assume the forward and the backward channels as fading channels, but the nodes are not interested in both these channel gains. Thus only partial information is needed to decide upon the feedback and the power levels. In our scheme, the receiver learns the channel state information which is a continuous random variable unlike learning a discrete message in [22]. Moreover in our scheme, the feedback channel has noise and is quantized in the case of FDD protocols unlike [22] where the feedback is a real number received noiselessly at the transmitter. Hence the basic idea of successive refinement in the two schemes is similar, but the schemes differ significantly.

Multiple round protocols are also used in ARQ systems [13]. In [13], the receiver sends an acknowledgement confirming if the data can be decoded in each round. However, it was assumed that the receiver knows the channel state information perfectly and the feedback is perfect. In this paper, the receiver obtains the channel state information by training and the feedback is sent over a noisy fading channel. In this paper, we consider the achievable diversity multiplexing tradeoff when the transmitter and receiver "conference" about the channel, which allows the transmitter to perform a more precise power control and achieve better performance. The conferencing assumes no Genie-knowledge at any point, and can be viewed as a consensus problem over noisy channels.

The rest of the paper is organized as follows. We formulate the problem in Section II. Section III and IV describes the known and the new results for the FDD and TDD systems respectively. Section V concludes the paper.

## II. PROBLEM FORMULATION

*A. Two-way Channel Model*

Consider a multiple input-output channel with the transmitting node denoted by T and receiving node denoted by R. We will assume that there are $m$ transmit antennas at the source node and $n$ receive antennas at the destination node, such that the input-output relation is given by

$$\mathsf{T} \to \mathsf{R} : Y = HX + W, \tag{1}$$

where the elements of $H$ and $W$ are assumed to be i.i.d. with complex normal distribution of zero mean and unit variance, $CN(0,1)$. The matrices $Y, H, X$ and $W$ are of dimension $n \times T_{coh}, n \times m, m \times T_{coh}$ and $n \times T_{coh}$, respectively. We assume that $T_{coh}$ is coherence interval such that the channel $H$ is fixed during a fading block of $T_{coh}$ consecutive channel uses, and statistically independent from one block to another. We further assume that $T_{coh}$ is finite and do not scale with SNR. The transmitter is assumed to be power-limited, such that the long-term power is upper bounded, i.e, $\frac{1}{T_{coh}}\text{trace}(\mathbb{E}\left[XX^\dagger\right]) \leq \mathsf{SNR}$.

We consider both frequency-division and time-division duplex models for the feedback channel. In both cases, we assume that the same multiple antennas at the transmitter and receiver are available to send feedback, in a half-duplex manner. For the feedback path, the receiver will act as a transmitter and the transmitter as a receiver. As a result, the feedback source (which is destination for data bits) will have $n$ transmit antennas and feedback destination (which is source of data bits) will be assumed to have $m$ receive antennas. Furthermore, a block fading channel model is assumed for the feedback channel,

$$\mathsf{R} \to \mathsf{T} : Y_f = H_f X_f + W_f, \tag{2}$$

where $H_f$ is the MIMO fading channel for the feedback link, and the $W_f$ is the additive noise at the receiver of the feedback; both are assumed to have i.i.d. $CN(0,1)$ elements. The feedback transmissions are also assumed to be power-limited with a long-term power constraint given by $\frac{1}{T_{coh}}\text{trace}(\mathbb{E}\left[X_f X_f^\dagger\right]) \leq \mathsf{SNR}_f$. Without loss of generality, we will assume the case where the transmitter and receiver have *symmetric resources*, such that $\mathsf{SNR} = \mathsf{SNR}_f$. All our results can be easily generalized to the case of asymmetric resource usage.

For the case of FDD systems, $H$ and $H_f$ are statistically independent. On the other hand, for the case of TDD, we assume that the $H$ and $H_f$ are perfectly correlated within one coherence interval, and adopt a phase-symmetric two-way channel model with $H_f = H^T$ [4, 5].

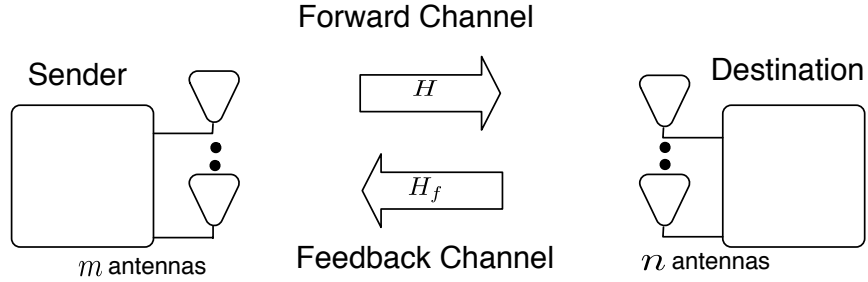

Fig. 2. Two-way fading channel, where the forward and feedback channels use the same antennas.

### B. Multi-round Protocols to Acquire Channel State

The two-way channel model allows the transmitter and receiver to conduct multiple rounds of information exchange. In this paper, we will consider coding strategies where the transmitter-receiver perform a multi-round exchange to estimate the channel $H$, before sending the data in order to maximize system diversity order. A *round* is defined to consist of two transactions: a transmission sent from node T to node R and a return transmission from node R to node T. We also define *half-round*, where only one node sends a transmission without receiving a transmission in exchange.

Figure 3 depicts the input-output signals and the temporal dependence between them. In round $i$, transmission from node T is denoted as $X_i\left(X^{i-1}, Y_f^{i-1}\right)$ where $X^{i-1} = \{X_1, X_2, \ldots, X_{i-1}\}$ are all the previous inputs and $Y_f^{i-1} = \{Y_{f,1}, Y_{f,2}, \ldots, Y_{f,i-1}\}$ are all the previous outputs of the feedback channel. Analogously, we define feedback channel input $X_{f,i}\left(X_f^{i-1}, Y^i\right)$. Since both T and R are assumed to be half-duplex and the multi-round protocol is always initiated by the transmitter T, the signal $X_{f,i}$ from node R can depend on the last round of received signal $Y_i$.

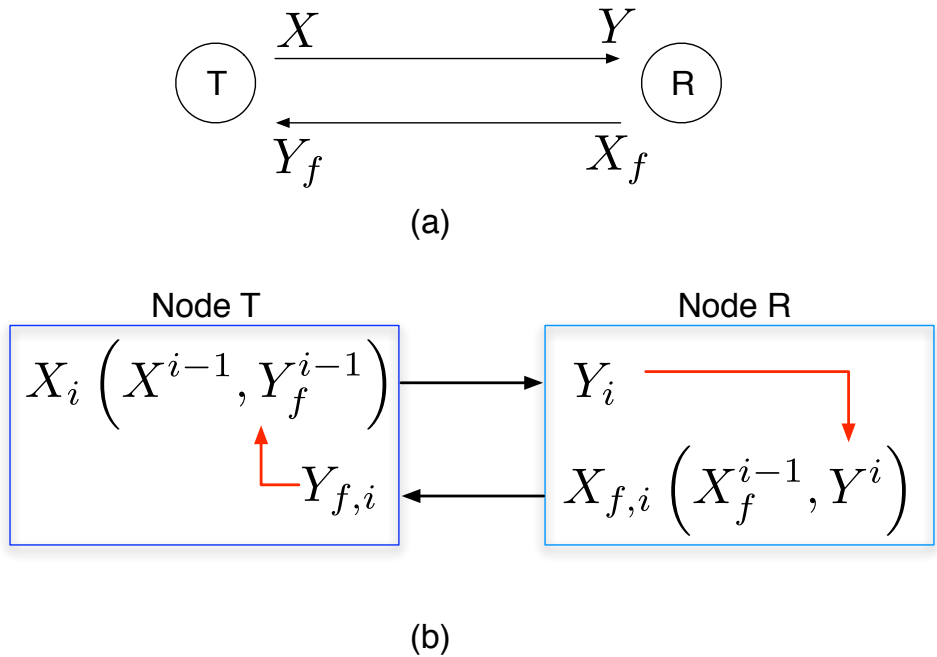

Fig. 3. Multi-round protocols: (a) Input-output relations and (b) temporal dependence of signals.

### C. Diversity-Multiplexing Tradeoff

In this paper, we will focus on the asymptotic regime of large SNR. Thus, we adopt the notation of [25] to denote $\doteq$ to represent exponential equality as $a \doteq b$ if $\lim_{\mathsf{SNR}\to\infty} \log(a)/\log(\mathsf{SNR}) = \lim_{\mathsf{SNR}\to\infty} \log(b)/\log(\mathsf{SNR})$. However, $a \doteq 0$ would mean that $a$ decays faster than any polynomial in SNR. In this paper, we are only concerned with probability decaying polynomially with SNR and hence the probability of events that decay faster than polynomial decay (example, exponential decay) will not enter the calculations. We similarly use $\dot{<}, \dot{>}, \dot{\leq}, \dot{\geq}$ to denote exponential inequalities.

We will only consider the case when the codeword $X$ spans a single fading block. Based on the transmitter channel knowledge $G$, the transmitted codeword is chosen from the codebook $C_G = \{X_G(1), X_G(2), \cdots, X_G(2^{RT_{coh}})\}$, where $R$ is the rate of the codebook. All $X_G(k)$'s are matrices of size $m \times T_{coh}$. In this paper, we will only consider single rate transmission where the rate of the codebooks does not depend on the transmitter knowledge, and the codebooks are derived from the same base codebook $C$ by power scaling of the codewords. In other words $C_G = \sqrt{P_G}C$, where the product implies that each element of every codeword is multiplied by $\sqrt{P_G}$ where each codeword $X(k) \in C$ has unit power. Thus, $P_G$ is the power of the transmitted codewords. Recall that there is an average transmit power constraint, such that $\mathbb{E}(P_G) \leq \mathsf{SNR}$. All coding strategies in this paper assume Gaussian input distribution.

Since our focus is on the delay-limited regime of single codewords, we will use outage as our metric. Outage is defined as the event when the mutual information of the channel given instantaneous realization of the estimate $H_R$ of the channel $H$ at the receiver, $I(X;Y|H_R)$ is less than the desired rate $R$. Let $\Pi(\mathcal{O})$ denote the probability of outage, where $\mathcal{O}$ is the set of all the channels where the transmitted rate $R$ is less than the maximum supportable rate $I(X;Y|H_R)$. The system is said to have diversity order of $d$ if $\Pi(\mathcal{O}) \doteq \mathsf{SNR}^{-d}$. Note that all the index mappings, codebooks, rates, powers are dependent on the average signal to noise ration, SNR. Specifically, the dependence of rate $R$ on SNR is explicitly given by $R = r \log \mathsf{SNR}$, where $r$ is labeled as the multiplexing gain. The diversity-multiplexing tradeoff is then described as the maximum diversity order $d(r)$ that can be achieved for a given multiplexing gain $r$.

If $H_R = H$, $I(X;Y|H) = \log \det \left(I + \frac{P_G}{m}HQH^\dagger\right)$ is the mutual information of a point-to-point link with $m$ transmit and $n$ receive antennas, transmit signal to noise ratio $P_G$ and input distribution Gaussian with covariance matrix $Q$ [25]. The dependence of the index at the transmitter is made explicit by writing the transmit SNR as a function of transmitter channel knowledge $G$. If $H_R \neq H$, we will consider an achievable strategy considering the estimation error as noise as in [26] and hence $2^{I(X;Y|H_R)} \dot{\geq} \det\left(I + \frac{P_G H_R H_R^\dagger}{(1+P_G \text{trace}(\mathbb{E}[(H-H_R)(H-H_R)^\dagger]))}\right)$ [26].

For the case of perfect receiver information and no transmitter information, let the codebooks of rate $R \doteq r \log \mathsf{SNR}$ and power $P \doteq \mathsf{SNR}^p$ are used. The outage event is defined as $\mathcal{O}(R,P) = \{H : \left(I + \frac{P}{m}HQH^\dagger\right) < R\}$, the probability of which is $\Pi(\mathcal{O}(R,P)) \doteq \mathsf{SNR}^{-G(r,p)}$ thus giving the diversity-multiplexing tradeoff of $d = G(r,p)$ [11], where

$$G(r,p) \triangleq \inf_{\alpha_1^{\min(m,n)} \in A(r,p)} \sum_{i=1}^{\min(m,n)} (2i - 1 + \max(m,n) - \min(m,n))\,\alpha_i,$$

and

$$A(r,p) \triangleq \{\alpha_1^{\min(m,n)} | \alpha_1 \geq \ldots \alpha_{\min(m,n)} \geq 0, \sum_{i=0}^{\min(m,n)} (p - \alpha_i)^+ < r\}.$$

The function $G(r,p)$ defines a piecewise linear curve connecting the points $(r, G(r,p)) = (kp, p(m-k)(n-k))$, $k = 0, 1, \ldots, \min(m,n)$ for fixed $m$, $n$ and $p > 0$. We define $G_u(r)$ recursively as follows. Let $G_0(r) = 0$, and $G_u(r) = G(r, 1 + G_{u-1}(r))$ for $u \geq 1$.

**Remark 1.** Consider a MIMO system with $m$ transmit and $n$ receive antennas with perfect channel state information at the receiver. The diversity multiplexing tradeoff with no CSIT is $d_{\mathrm{CSIR}} = G(r, 1) = G_1(r)$ [25]. Further, the diversity multiplexing tradeoff of $d_{\mathrm{CSIRT_q}} = G_K(r)$ can be achieved with $K$ levels (or equivalently $b = \log_2(K)$ bits) of quantized feedback from the receiver about the channel [8].

## III. FDD PROTOCOLS: ITERATIVE QUANTIZATION

In this section, we extend our 1.5 round protocol of [6] in two directions. First, we show that for non-zero multiplexing gains, more than 1-bit of information about the channel can in fact be extracted and sent reliably across a noisy feedback channel. Furthermore, the number of bits of information which can be extracted depends on the multiplexing gain. Thus a higher diversity order can be achieved than the proposed techniques in [6] for $r > 0$. Second, we construct a multi-round protocol, which can extract more bits about the channel at *all* multiplexing gains, including zero multiplexing point. The protocol iteratively refines the channel information at both ends, and also keeps the two nodes aligned in their knowledge about the channel estimate.

### A. Prior Results: 1-bit about the channel in 1.5 rounds

The diversity-multiplexing tradeoff for the case of receiver with perfect information and noiseless quantized information has been extensively studied in [2, 8, 9]. In this subsection, we will review the results of [6] where imperfect training and noisy feedback was considered for FDD systems, described in the new terminology of Section II.

1) **Round 1 (forward)**: T sends a training at power SNR, $X_1 = \sqrt{\text{SNR}}\,\beta$, where $\beta$ is known fixed sequence. The signal received at R is $Y_1 = HX_1 + W$, which is used to form an MMSE estimate $\widehat{H}_1$.
   **Round 1 (reverse)**: The feedback from R is a 1-bit quantization of the channel estimate $q_1 : \widehat{H}_1 \mapsto \{0,1\}$, of the estimated channel such that
   $$q_1(\widehat{H}_1) = \begin{cases} 0, & \widehat{H}_1 \in \widehat{\mathcal{O}}_1^c \\ 1, & \text{otherwise} \end{cases}, \tag{3}$$
   where $\widehat{\mathcal{O}}_1 = \{\widehat{H}_1 : \det(I + \widehat{H}_1 \widehat{H}_1^\dagger \text{SNR}) < \text{SNR}^{r+\epsilon}\}$ The quantized bit is sent as follows:
   $$X_{f,1}(Y_1) = \begin{cases} \sqrt{\text{SNR}^0}\,\beta_f, & q_1(\widehat{H}_1) = 0 \\ \sqrt{\text{SNR}^{1+G(r+\epsilon,1)}}\,\beta_f, & q_1(\widehat{H}_1) = 1 \end{cases}, \tag{4}$$
   where $\beta_f$ is a known sequence. The signal received at node T is $Y_{f,1} = H_f X_{f,1} + W_f$.
2) **Round 2 (forward only)**: Node T decodes the bit of information about the channel from $Y_{f,1}$, labeled as $\widehat{q}_1$ (found by received power estimate with thresholds of $\text{SNR}^{\delta/mn}$ for some $\delta << \epsilon$), and forms a power controlled signal
   $$X_2(Y_{f,1}) = \begin{cases} \sqrt{\text{SNR}^1}\,x, & \widehat{q}_1 = 0 \\ \sqrt{\text{SNR}^{1+G(r+\epsilon,1)}}\,x, & \widehat{q}_1 = 1 \end{cases}, \tag{5}$$
   where $x = \begin{bmatrix} \beta & c \end{bmatrix}$, where $c$ is the data codeword concatenated to the training signal $\beta$.

**Theorem 1** ([6]). *For $\epsilon < r < m - \epsilon$, the above strategy achieves the diversity multiplexing tradeoff of $d_{1.5,2}(r) = G(r, 1 + G(r,1))$ by choosing $\epsilon$ arbitrarily close to 0.*

**Remark 2.** Note that the signals in the protocol depend on the last channel output; in particular $X_{f,1}(Y_1)$ and $X_2(Y_{f,1})$. In the multi-round protocol discussed in Section III-C, we will consider the generalized feedback structures, which will depend on all previous channel outputs, as described in Section II-B.

**Remark 3.** Unlike the previous work in [4, 26], we do not account for the resources spent in channel training and feedback in this paper. Resource accounting can be performed using the procedure developed in [4], by scaling the multiplexing $r$ appropriately. More precisely, the multiplexing $r$ should be replaced by $rT/(T - 2m - 1)$. The resource accounting multiplier assumes that the feedback requires one channel use and the training requires $m$ channel uses. Further, the number of antennas would need to be optimized for each multiplexing gain, as in [4, 26].

A quick note on the notation $d_{p,K}(r)$ for diversity-multiplexing tradeoff: the first subscript $p$ refers to the number of rounds in the protocol and the second subscript $K$ refers to the number of levels communicated by the receiver in each round. All FDD protocols discussed in this paper rely on receiver sending quantized information about the measured channel back to the transmitter. In the TDD protocols, the channel symmetry will be exploited and transmitter will also convey channel information to the receiver. Further in TDD protocols, a power-controlled training symbol may be fed back from the receiver which takes $m$ channel uses and is equivalent to sending soft information about the channel, instead of quantized bits like in FDD protocols. Thus, the second subscript $K$ will be suppressed and $d_p(r)$ will be used to denote diversity order of TDD protocols.

*Example 1 (1.5 round protocol, 1 bit feedback)*: Consider a SISO system where the receiver now does not know the value of channel estimate $H$, but only an estimate $\widehat{H}_1$. The above strategy reduces to the receiver deciding if $\widehat{H} \in \widehat{\mathcal{O}}_1$, where
$$\widehat{\mathcal{O}}_1 = \left\{ \widehat{H}_1 : \log\left(1 + |\widehat{H}_1|^2 \text{SNR}\right) < R \right\}. \tag{6}$$

Let $\alpha$ be the negative SNR exponent of $|H|^2$ while $\widehat{\alpha}$ be the negative SNR exponent of $|\widehat{H}_1|^2$. Note that $\widehat{\mathcal{O}}_1$ represent the event $\widehat{\alpha} > 1 - r$. Since $\alpha = \widehat{\alpha}$ with probability 1 when $\widehat{\alpha} < 1$ (For more details, the reader is referred to Appendix A-1.), this decision from the receiver is reliable. Further since $\Pi(\widehat{\mathcal{O}}_1) = \text{SNR}^{-(1-r)}$, the receiver uses a power level of $\text{SNR}^{2-r}$ to communicate $\widehat{\mathcal{O}}_1$, while $\text{SNR}^0$ to communicate the possibility of $\widehat{\mathcal{O}}_1^c$. This is a special case of using power-controlled feedback; the reader is referred to Appendix A-2. If the transmitter gets $\widehat{\mathcal{O}}_1$, it transmits with power $\text{SNR}^{2-r}$, else uses SNR. Hence, the two error events are that the transmitter made an error in decoding which happens with probability $\text{SNR}^{-2+\epsilon}$ while the second is that the power of $\text{SNR}^{2-r}$ is not sufficient which happens with probability $\text{SNR}^{-2(1-r)}$. Thus, diversity of $2(1-r)$ can be achieved. ∎

### B. First Extension: $\log_2(K)$ bits about the channel in 1.5 rounds

In this subsection, we will extend the results of Theorem 1 to more than 1 bit of feedback for non-zero multiplexing gain. The key observation is that while no more than 1-bit can be extracted for zero multiplexing gain [6], the channel estimate $\widehat{H}_1$ has enough resolution to derive more bits about for non-zero multiplexing gains, $r > 0$. The modified protocol can be described as follows.

1) **Round 1 (forward)**: T sends a training at power SNR, $X_1 = \sqrt{\mathsf{SNR}}\beta$, which is used by the receiver to form an MMSE estimate $\widehat{H}_1$.
   **Round 1 (reverse)**: Let
   $$\widehat{\mathcal{O}}_{K-1} = \left\{\widehat{H}_1 : \det(I + \widehat{H}_1\widehat{H}_1^\dagger \mathsf{SNR}^{1+\min(c_M, G_{K-2}(r+\epsilon))}) < \mathsf{SNR}^{r+\epsilon}\right\}, \tag{7}$$
   where $c_M = \left(\frac{r-m+1}{m}\right)^+$, where the notation $(x)^+$ denotes $\max(x,0)$. Further, for any $u \in [0, K-2]$, let $\widehat{\mathcal{O}}_u$ be defined as
   $$\widehat{\mathcal{O}}_u = \widehat{\mathcal{O}}_{K-1}^c \bigcap \left\{\bigcap_{j=0}^{u-1} \widehat{\mathcal{O}}_j^c\right\} \bigcap \left\{\widehat{H}_1 : \det(I + \widehat{H}_1\widehat{H}_1^\dagger \mathsf{SNR}^{1+\min(c_M, G_u(r+\epsilon))}) \geq \mathsf{SNR}^{r+\epsilon}\right\}. \tag{8}$$
   The feedback from R is a $K$ level quantization, $q_1 : \widehat{H}_1 \mapsto \{0, 1, \ldots, K-1\}$ computed as follows,
   $$q_1(\widehat{H}_1) = u \text{ if } \widehat{H}_1 \in \widehat{\mathcal{O}}_u. \tag{9}$$
   Let $p_u = \min(G_u(r+\epsilon), c_M)$. The quantized signal $q_1(\widehat{H}_1)$ is sent as
   $$X_{f,1}(Y_1) = \begin{cases} \sqrt{\mathsf{SNR}^{(c_M - G_u(r+\epsilon))^+}}\beta_f, & q_1(\widehat{H}_1) = u < K-1 \\ \sqrt{\mathsf{SNR}^{1+G(r+\epsilon, 1+p_{K-2})}}\beta_f, & q_1(\widehat{H}_1) = K-1 \end{cases}. \tag{10}$$
   The signal received at node T is $Y_{f,1} = H_f X_{f,1} + W_f$.

2) **Round 2 (forward only)**: Node T decodes the $K$ level information about the channel from $Y_{f,1}$, labeled as $\widehat{q}_1$ (found by received power estimate with thresholds of $\mathsf{SNR}^{(c_M - G_u(r+\epsilon))^+ + \delta/mn}$ for some $\delta \ll \epsilon$), and forms a power controlled signal
   $$X_2(Y_{f,1}) = \begin{cases} \sqrt{\mathsf{SNR}^{1+(p_u - \delta)^+}} x, & \widehat{q}_1 = u < K-1 \\ \sqrt{\mathsf{SNR}^{1+\min(G_{K-1}(r+\epsilon), G(r+\epsilon, 1+c_M))}} x, & \widehat{q}_1 = K-1 \end{cases}, \tag{11}$$
   where $x = \begin{bmatrix} \beta & c \end{bmatrix}$, where $\beta$ is known training signal as above and $c$ is the data codeword.

**Theorem 2.** *For $\epsilon < r < m - \epsilon$, the above strategy achieves the diversity multiplexing tradeoff of $d_{1.5,K}(r) = \min\{G(r, 1 + G(r, 1 + c_M)), G_K(r)\}$ by choosing $\epsilon$ arbitrarily close to 0.*

*Proof:* The proof is provided in Appendix B. ∎

We first note that for $r > m-1$ and $K > 2$, the diversity order $d_{1.5,K}(r)$ achieved in Theorem 2 is greater than the diversity order $d_{1.5,1}(r)$ achieved in Theorem 1. For $r > m-1$, a finer power control is possible in the region where the channel condition is good, i.e, where the channel estimate dominates the noise floor. In contrast, when $r \to 0$, the channel is not resolvable beyond one bit of information. The increased channel resolvability at higher multiplexing gains is the main reason for increased diversity order. The following example makes the above intuition explicit.

**Example 2 (1.5 round protocol, 3 level feedback, $r = 1/6$):** Consider a SISO system where the receiver only has a channel estimate, $\widehat{H}_1$. The above strategy reduces to the receiver deciding if $\widehat{H}_1 \in \widehat{\mathcal{O}}_i$, where
$$\widehat{\mathcal{O}}_2 = \left\{\widehat{H}_1 : \left(1 + |\widehat{H}_1|^2 \mathsf{SNR}^{7/6}\right) \leq \mathsf{SNR}^{1/6}\right\} \tag{12}$$
$$\widehat{\mathcal{O}}_1 = \widehat{\mathcal{O}}_2^c \cap \left\{\widehat{H}_1 : \left(1 + |\widehat{H}_1|^2 \mathsf{SNR}\right) \leq \mathsf{SNR}^{1/6}\right\} \tag{13}$$
$$\widehat{\mathcal{O}}_0 = \widehat{\mathcal{O}}_1^c \cap \widehat{\mathcal{O}}_2^c. \tag{14}$$

Let $\alpha$ be the negative SNR exponent of $|H|^2$ while $\widehat{\alpha}$ be the negative SNR exponent of $|\widehat{H}_1|^2$. Note that $\widehat{\mathcal{O}}_1$ represent the event $5/6 \leq \widehat{\alpha} < 1$ and $\widehat{\mathcal{O}}_2$ represent the event $\widehat{\alpha} \geq 1$. Since $\alpha = \widehat{\alpha}$ with probability 1 when $\widehat{\alpha} < 1$, this decision from the receiver is reliable. The three events $\widehat{\mathcal{O}}_0$, $\widehat{\mathcal{O}}_1$ and $\widehat{\mathcal{O}}_2$ are transmitted from the receiver at powers of $\mathsf{SNR}^{1/6}$, $\mathsf{SNR}^0$ and $\mathsf{SNR}^2$ respectively. The power levels used by the transmitter in these three events are $\mathsf{SNR}$, $\mathsf{SNR}^{7/6}$ and $\mathsf{SNR}^2$ respectively. It is easy to see that the power constraints are asymptotically satisfied. Outage occurs when less power is used due to error in the feedback or when the power of $\mathsf{SNR}^2$ was not sufficient. The event that less power is used due to feedback error happens with probability $\mathsf{SNR}^{-(2-1/6)+\epsilon}$ and the event that power of $\mathsf{SNR}^2$ is not sufficient to avoid outage happens with probability $\mathsf{SNR}^{-(2-1/6)}$. Hence, a diversity of $2 - 1/6 = 2 - r$ can be achieved. Note that this is larger than $2 - 2r$ which is achieved with 1 bit of feedback. The increase with the additional level of the feedback is because $\alpha = \widehat{\alpha}$ (with probability 1) as long as $\widehat{\alpha} < 1$ which gives perfect resolution for $\alpha$ from the estimate as long as $\widehat{\alpha} < 1$. The first feedback decides whether $\alpha \geq 5/6$ and since there is a gap between $5/6$ and $1$ in which the channel can be resolved perfectly, the next feedback level resolved whether the channel satisfies $5/6 \leq \alpha < 1$. However at multiplexing $r \to 0$, no more than 1-bit of information could be

derived from the estimate since the channel cannot be resolved when $\widehat{\alpha} \geq 1$. ∎

In fact, the number of useful bits which can be derived about the channel can be linked to the multiplexing gain as shown in the following corollary.

**Corollary 1** (Bits About the Channel). *The maximum number of bits $b$ about the channel, for sake of power control, which can be derived from the channel estimate $\widehat{H}_1$ as a function of multiplexing gain $r$ is given as follows.*

1) *SISO:* $\log_2(K+2)$ *noiseless bits about the channel can be derived if the multiplexing gain* $r \in \left(\frac{K-1}{K}, \frac{K}{K+1}\right]$.
2) *MIMO: For $r \leq m-1$, one noiseless bit can be derived. For $m-1 < r < m$, $\log(K+3)$ bits can be derived, where $K = \max\{u : u \geq 0, c_M > G_u(r)\}$.*

The final corollary of this section captures the performance of $d_{1.5,K}(r)$ as the number of levels increases. We note that due to error in channel estimation and the feedback channel, the diversity order $d_{1.5,K}(r)$ does not increase unboundedly with $K$ for a given $r$.

**Corollary 2.** *As $K \to \infty$, the above diversity multiplexing tradeoff reduces to $G(r, 1 + G(r, 1 + c_M))$ for any $0 < r < \min(m, n)$. This is because $G_K(r)$ is a monotonic increasing function with $K$ and goes to $\infty$ as $K \to \infty$.*

For the special case of $m = 1$, $c_M = r$ and $\lim_{K \to \infty} d_K(r) = n^2 + n(1-r)$. Kim *et al.* considered a model of imperfect channel estimate at the transmitter in [10] where the transmitter knows the correlated channel estimate with a certain correlation while the receiver knows the channel state information perfectly. However in our case, the receiver knows a correlated version of the channel state which is shared with the transmitter via noisy quantized feedback. Even though the two cases are different, in the special case of $m = 1$, both the models yield the same diversity multiplexing tradeoff.

### C. Iterative Quantization: $\log_2(K)$ bits in $(K-1).5$ Rounds

In the previous two sections, the protocols used only 1.5 rounds. As a result, for $r \to 0$, no more than 1-bit about the channel could be extracted. In this section, we show how multiple rounds can be used by the transmitter and receiver to iteratively refine the knowledge about the channel $H$ beyond one bit for *all* multiplexing gains, including $r = 0$. In each round, the receiver sends a binary signal back to the transmitter, providing an additional level of channel information. To perform the iterative quantization, both transmitter and receiver occasionally and opportunistically use large transmit power in some blocks. However, the use of large power is performed very rarely, allowing both nodes to stay within their prescribed average power constraints over the long-term.

For sake of clarity, we will state the *multi-round iterative quantization protocol* when only binary messages are conveyed by the receiver in each round. Thus there is 1-bit quantization by the receiver in each round. All our results can be extended to feedback messages belonging to a larger alphabet in each round (like in Section III-B).

1) **Round 1 (forward)**: T sends a training at power SNR, $X_1 = \sqrt{\text{SNR}}\beta$, where $\beta$ is a known fixed sequence. The signal received at R is $Y_1 = HX_1 + W$, which is used to form an MMSE estimate $\widehat{H}_1$. The feedback from R is a binary quantization, $q_1(\widehat{H}_1)$ such that

$$q_1(\widehat{H}_1) = \begin{cases} 0, & \text{if } \widehat{H}_1 \in \widehat{\mathcal{O}}_1^c \\ 1, & \text{otherwise} \end{cases}, \quad (15)$$

where $\widehat{\mathcal{O}}_1 = \{\widehat{H}_1 : \det(I + \widehat{H}_1 \widehat{H}_1^\dagger \text{SNR}) < \text{SNR}^{r+\epsilon}\}$.

**Round 1 (reverse)**: The quantized channel $q_1(\widehat{H}_1)$ is modulated as follows:

$$X_{f,1}(Y_1) = \begin{cases} \sqrt{\text{SNR}}\beta_f, & q_1(\widehat{H}_1) = 0 \\ 0 \cdot \beta_f, & q_1(\widehat{H}_1) = 1 \end{cases}, \quad (16)$$

where $\beta_f$ is a known sequence. The signal received at node T is $Y_{f,1} = H_f X_{f,1} + W_f$. The transmitter T estimates $q_1$ as $\widehat{q}_1$ using a MAP detector for power level with a threshold of $\text{SNR}^{\epsilon/mn}$.

At the end of round $i-1$, the transmitter forms an estimate $\widehat{q}_{i-1}$ of the quantization performed at the receiver, which is denoted by the function $q_{i-1}(\cdot)$. To state the following recursion, we assume $\widehat{q}_0 = 1$. The estimate $\widehat{q}_{i-1}$ is used in the $i^{th}$ round by the transmitter as described below.

2) **Round $i \in \{2, \ldots, K-1\}$ (forward)**: The transmitter trains the receiver with $X_i = \sqrt{P_i}\beta$ where the power level $P_i$ chosen as follows

$$P_i = \begin{cases} 0, & \text{if } \widehat{q}_{i-1} = 0 \\ \text{SNR}^{1+G_{i-1}(r+\epsilon)}, & \text{otherwise} \end{cases}. \quad (17)$$

The receiver's actions can be described as follows.

- If any of $q_u(\widehat{H}_u, q^{u-1}) = 0$ for $u \leq i-1$, then the receiver performs a MAP power estimation to check if the transmitter sent a training at power level of $0$ or $\mathsf{SNR}^{1+G_{i-1}(r+\epsilon)}$. The associated MAP threshold is $\mathsf{SNR}^\epsilon$. If power estimate $\geq \mathsf{SNR}^\epsilon$, $q_i(\widehat{H}_i, q^{i-1}) = 0$, else $q_i(\widehat{H}_i, q^{i-1}) = 1$.
- If $q_u(\widehat{H}_u, q^{u-1}) = 1$ for all $u \leq i-1$, the receiver estimates the channel as $\widehat{H}_i$ assuming that the training power is $\mathsf{SNR}^{1+G_{i-1}(r+\epsilon)}$ and bases the feedback signal based on the estimate as follows

$$q_i(\widehat{H}_i, q^{i-1}) = \begin{cases} 0, & \text{if } \widehat{H}_i \in \widehat{\mathcal{O}}_i^c \\ 1, & \text{otherwise} \end{cases}, \quad (18)$$

where $\widehat{\mathcal{O}}_i = \{\widehat{H}_i : \det(I + \widehat{H}_i \widehat{H}_i^\dagger \mathsf{SNR}^{1+G_u(r+\epsilon)}) < \mathsf{SNR}^{r+\epsilon}\}$.

3) **Round** $i \in \{2, \ldots, K-1\}$ (**reverse**): The receiver sends the signal

$$X_{f,i} = \begin{cases} \sqrt{\mathsf{SNR}^{1+G_{i-1}(r+\epsilon)}} \, \beta_f, & q_i(\widehat{H}_i, q^{i-1}) = 0 \\ 0 \cdot \beta_f, & \text{otherwise} \end{cases}, \quad (19)$$

where $\beta_f$ is a known sequence. The signal received at node $\mathsf{T}$ is $Y_{f,i} = H_f X_{f,i} + W_f$. If $\widehat{q}_{i-1} = 0$, then $\widehat{q}_i = 0$. That is, if the transmitter had concluded that it was a "good" channel after round $i-1$, then it does not attempt to re-estimate the feedback signal. Otherwise, the transmitter decodes the power level by a MAP detection with a threshold of $\mathsf{SNR}^{\epsilon/mn}$ to get an estimate of $q_i$ to get $\widehat{q}_i$.

4) **Round** $K$ (**forward only**): The transmitter then trains the receiver and then sends the data both at a power level of $\mathsf{SNR}^{1+G_u(r+\epsilon)}$ where $u = \min\{\{v \in [1, K-1] : \widehat{q}_v = 0\} \bigcup K\} - 1$.

**Theorem 3.** *The iterative quantization protocol, described by above Steps 1-5, achieves a diversity order of* $d_{(K-1).5,1} = G_K(r)$.

*Proof:* The proof is provided in Appendix C. ∎

Note that this diversity is same as that with $K$ levels of perfect feedback when the receiver knows perfect channel state information as given in [8]. Thus $(K-1).5$ rounds allow the transmitter and receiver to *agree* upon $\log_2(K)$ bits about the channel. Note that the bits about the channel $H$ relate to the outage regions, and hence the Voronoi regions are concentric spheres, all of which are centered at origin. In Section III-D, we will provide further intuition to the above result. The following example serves as the stepping stone to the general discussion.

**Example 3 (2.5 round iterative quantization, 1 bit feedback per round)**: Consider a SISO system where the receiver now does not know the value of channel estimate $H$, but only an estimate $\widehat{H}_1$. The above strategy reduces to the receiver deciding if $\widehat{H}_1 \in \widehat{\mathcal{O}}_1$, where

$$\widehat{\mathcal{O}}_1 = \left\{\widehat{H}_1 : \log\left(1 + |\widehat{H}_1|^2 \mathsf{SNR}\right) < R = r \log(\mathsf{SNR})\right\}. \quad (20)$$

Let $\alpha$ be the negative SNR exponent of $|H|^2$ while $\widehat{\alpha}$ be the negative SNR exponent of $|\widehat{H}_1|^2$. Note that $\widehat{\mathcal{O}}_1$ represent the event $\widehat{\alpha} > 1 - r$. Since $\alpha = \widehat{\alpha}$ with probability 1 when $\widehat{\alpha} < 1$, this decision from the receiver is reliable. The receiver uses a power level of $\mathsf{SNR}^0$ to communicate $\widehat{\mathcal{O}}_1$ and $\mathsf{SNR}$ to communicate the possibility of $\widehat{\mathcal{O}}_1^c$. If the transmitter receives $\widehat{\mathcal{O}}_1^c$, then this was the only possibility of being transmitted (ignoring the events with probability $\doteq 0$). However, if the transmitter receives $\widehat{\mathcal{O}}_1$, $\widehat{\mathcal{O}}_1^c$ could still have been transmitted and the error happened with probability $\mathsf{SNR}^{-1}$.

In the second round, the transmitter trains at a power of $\mathsf{SNR}^{2-r}$ only if it receives $\widehat{\mathcal{O}}_1$. If the receiver earlier sent $\widehat{\mathcal{O}}_1^c$, the channel was good. Hence, a MAP detection is done to estimate if the transmitter made a mistake. The error made by the transmitter would be perfectly known at the receiver in this case. If the receiver earlier decided that the channel was in $\widehat{\mathcal{O}}_1$, it now estimates the channel $\widehat{H}_2$ and decides if $\widehat{H}_2 \in \widehat{\mathcal{O}}_2$, where

$$\widehat{\mathcal{O}}_2 = \left\{\widehat{H}_2 : \log\left(1 + |\widehat{H}_2|^2 \mathsf{SNR}^{2-r}\right) < R = r \log(\mathsf{SNR})\right\}. \quad (21)$$

Note again that the decision is being made if $\widehat{\alpha}_i < p - r$ which will be correct as the estimate and the actual channel will have same exponent in this region. If the receiver sent $\widehat{\mathcal{O}}_2^c$ and there was no error at transmitter, or the receiver sent $\widehat{\mathcal{O}}_2$ and is in $\widehat{\mathcal{O}}_2$, the receiver uses a power level of $0$. However, in the remaining cases, power of $\mathsf{SNR}^{2-r}$ is used. If there is an error in feedback, higher power will be used and hence there is no outage associated with this error.

Now, the transmitter knows if $\mathsf{SNR}$ or $\mathsf{SNR}^{2-r}$ might be fine or not. Hence, uses the three power levels of $\mathsf{SNR}$, $\mathsf{SNR}^{2-r}$ and $\mathsf{SNR}^{3-2r}$ in these three cases. There is an outage when power level of $\mathsf{SNR}^{3-2r}$ is not sufficient to avoid outage, which happens with probability $\mathsf{SNR}^{-3(1-r)}$ giving a diversity of $3(1-r)$. ∎

*D. Discussion of Iterative Quantization*

Although, the diversity of $G_K(r)$ can be obtained with perfect channel state at the receiver and perfect quantized feedback of $K$ levels, achieving this gain when both the channel estimates and the feedback are noisy is a challenge. Since the receiver has to use training symbol to estimate the channel, it cannot resolve the channel gains which are smaller than the estimation error. Recall that power control gain results from using large instantaneous power in poor channel conditions. Since such poor channel states are rare, the transmitter has to use large power rarely and hence can still meet the average power constraint. However, it is precisely the identification of the poor channel states which is not possible due to error in channel estimation at the receiver.

Ideally, one should use a large power, say $\mathsf{SNR}^p, p > 1$, for the training signal, which is the sequence $\beta$ in the protocols described in Sections III-A, III-B and III-C. The large training power means that the error floor is reduced to $\mathsf{SNR}^{-p}$ and we can identify channels gains of the order of $\mathsf{SNR}^{-p}$ with high reliability. However, the use of $\mathsf{SNR}^p, p > 1$ in every frame will violate the average power constraint of $\mathsf{SNR}$.

The solution is to use large training power *only* when the channel gain is too small to be resolved using lesser training power. Thus, the probability of using large training power should be governed by the probability of channel event, and thus the weighted average of transmitter power will meet the average power constraint of $\mathsf{SNR}$. The above *on-demand use* of training power is precisely the main idea behind iterative quantization described in Section III-C.

The other key ingredient is power-controlled feedback, which skews the feedback errors in one direction. The receiver in each iteration communicates to the transmitter if the current power level will be enough to avoid outage. However some feedback errors are worse than others. Consider the case of binary feedback. If the channel is Good ($\log\det(I + HH^\dagger\mathsf{SNR}^p) \geq \tau$ for appropriate $p$ and $\tau$) but the transmitter decodes it as a Bad channel and uses larger power than needed, no outage occurs during data transmission. So confusing Good channels for Bad channels is a non-issue, especially due to additional round of power-controlled training. However, if a Bad channel is confused as a Good channel, then the transmitter will use lower power than needed and hence as a result, there will be a definite outage. Thus, confusing Bad channels with Good channels is the main contributor of outage due to feedback errors.

Thus the desired situation will be that $\mathrm{Prob}(\mathsf{Bad} \to \mathsf{Good})$ is as small as possible while ensuring that the probabilities of the events $\{\mathsf{Good}, \mathsf{Bad}\}$ as seen by transmitter is same as those seen by the receiver; this last constraint on event probabilities at transmitter ensures that average power consumed is $\mathsf{SNR}$. All of the power-controlled feedback designs proposed in previous sections skew the feedback power allocation to lower $\mathrm{Prob}(\mathsf{Bad} \to \mathsf{Good})$ at the expense of limiting the maximum power. This is because the probability of Good event is asymptotically much higher than the probability of Bad event and thus if higher power is used for the Good event, the maximum power would be limited by $\mathsf{SNR}/\Pi(\mathsf{Good})$. The Bad event is encoded with a codeword of zero power by the receiver on the feedback link, which is decoded by the transmitter using an energy-based estimator. The probability of confusing a zero power codeword with any codeword of power greater than $\mathsf{SNR}^\delta, \delta > 0$, i.e. $\mathrm{Prob}(\mathsf{Bad} \to \mathsf{Good})$, decays exponentially fast. In comparison, the probability of confusing a Good channel with a Bad due to decoding an SNR codeword as a zero power codeword decays polynomially fast.

The iterative quantization can be more easily understood by a pictorial view of the decision process in Example 3. In 2.5 rounds with 1-bit per feedback round, we can resolve channels into one of 3 regions depicted in Figure 5(c). In the first round (forward), the receiver can distinguish between the events $\widehat{\mathcal{O}}_1$ and $\overline{\widehat{\mathcal{O}}_1}$. This is communicated to the transmitter using a power-controlled feedback, where $q_1 = 1$ representing event $\overline{\widehat{\mathcal{O}}_1}$ is encoded with zero power codeword. As a result, if $q_1 = 1$, then $\widehat{q}_1$ will be 1. Now, consider the case when the actual event is $\widehat{\mathcal{O}}_1$. In this case, in the second round (forward), the transmitter sends a training at power $\mathsf{SNR}^2$ which allows the receiver to distinguish between $\widehat{\mathcal{O}}_1 \setminus \widehat{\mathcal{O}}_2$ and $\widehat{\mathcal{O}}_2$. Once again, if $q_2 = 1$, a zero-power codeword is used by the receiver, which implies that the transmitter concludes $\widehat{q}_2 = 1$ with exponentially small error probability. Hence, the transmitter knows $\widehat{q}_1 = 1$ inside $\widehat{\mathcal{O}}_1$ and $\widehat{q}_2 = 1$ inside $\widehat{\mathcal{O}}_2$ (with probability 1). Thus, we find that the transmitter has received an index that is the same or bigger than that transmitted by the receiver in all the scenarios. This indicates that there will be no error because of the transmitter receiving the index by feedback incorrectly since the transmitter will send at atleast the power that the receiver suggested (with probability 1). Since the feedback errors do not dominate and in each feedback and the receiver adds one level precision in each round, the diversity corresponding to $K$ levels of perfect feedback is attained with $(K-1).5$ rounds of iterative quantization protocol.

The key to iterative quantization is that the channel is resolved to a finer level only when needed. As and when the channel is below the channel estimation error floor, the transmitter sends a higher power training compared to previous rounds. The on-demand nature of the protocol ensures neither the transmitter, not the receiver exceed their power budget and use high powers only in rare cases.

## IV. TDD Protocols: Exploiting Channel Symmetry

In the FDD systems, the forward and feedback channels are independent. Thus, the feedback channel fading and errors had to be dealt independent of the forward channel, or in other words, information about one channel cannot be exploited for better encoding on the other channel. In contrast, the symmetry of TDD channels can be effectively exploited by using the knowledge of one channel for transmission on the other. We will show that this symmetry allows us to achieve higher diversity-multiplexing tradeoffs.

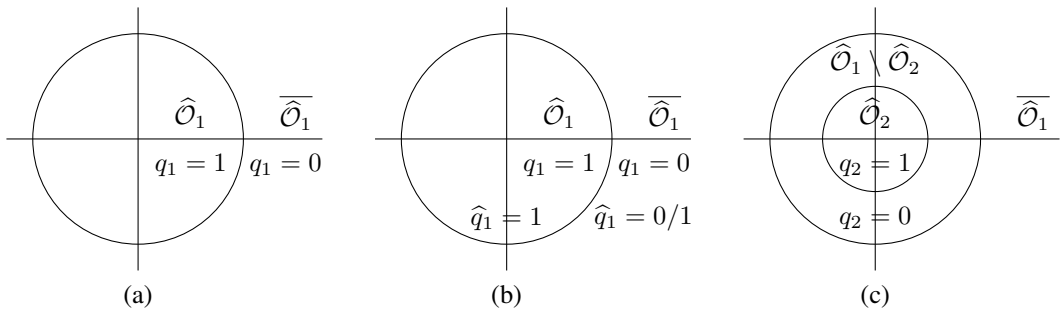

Fig. 4. Channel events (a) in the first round (forward), (b) in the first round (reverse) and (c) in the second round(forward).

The study of training based TDD systems from diversity multiplexing perspective was initiated in [4] for a MISO/SIMO system, where a receiver-initiated strategy was proposed and can be cast as a 1.5 round scheme. This was further extended to a MIMO system in [18], where it was shown that a diversity order upto $2mn$ (for zero multiplexing gain) was achievable.

In this section, we will generalize the prior result and propose a multi-round TDD protocol which iteratively refines the channel information at both ends, and also keeps the two nodes aligned in their knowledge about the channel estimate. Unlike the FDD iterative protocol, this protocol uses the fact that the errors in the forward and the feedback path will be correlated since both paths have same eigen-values due to the reciprocity of the channel. With this protocol, the diversity-multiplexing tradeoff is a straight line and dominates the diversity-multiplexing tradeoff for the FDD iterative quantization for all $0 < r < m$ with $(K-1).5$ rounds for any $K \geq 2$. For the special case of $K = 2$, the diversity multiplexing tradeoff of $mn(mn+1)-(m+n-1)r$ can be achieved using a power controlled strategy at the transmitter based on its own channel estimate. This power control is similar to that in [10] where the transmitter has a channel estimate with some error variance. While the model in [10] has no feedback channel, we can use the similar structure due to reciprocity of forward and feedback channels.

## A. Known Results:1.5 rounds

In this subsection, we review the result in [5, 18] for the achievable diversity multiplexing tradeoff for a MIMO channel. The authors considered a receiver initiated protocol, which in the terminology of this paper can be presented in 1.5 rounds in which the transmitter in the first round remains silent. The protocol is described as follows,

1) **Round 1 (forward):** The transmitter remains silent.
   **Round 1 (reverse):** The receiver sends a training at power SNR, $X_{f,1} = \sqrt{\text{SNR}}\beta$, where $\beta$ is known fixed sequence. The signal received at T is $Y_{f,1} = HX_{f,1} + W$, which is used to form an MMSE estimate $\widehat{H}_1$.
2) **Round 2 (forward only):** Let $\widehat{H}_1 = [\widehat{h}_1^T \cdots \widehat{h}_n^T]^T$ and $\lambda_j = \widehat{h}_j\widehat{h}_j^\dagger$. The transmitter trains the receiver and sends data using a power level of $\doteq \frac{\text{SNR}}{\min_{i=1}^n \lambda_i}$.

**Theorem 4** ([5]). *For $0 < r < 1$, the above protocol achieves a diversity multiplexing tradeoff of $mn(2-r)$. For $1 \leq r < m$, diversity order of $G(r, 1)$ is achievable.*

Thus, we see that the above protocol converts the MIMO system to a parallel MISO channels and uses the worst of the parallel channels to determine the power control. For $r < 1$, the power control mentioned in this protocol achieves better diversity as compared to the diversity achieved without without feedback. However for $r \geq 1$, the achievable diversity is same as that with no feedback system of [26] as feedback for each MISO channel improves the diversity only in the range $r < 1$.

In the case of FDD systems, if the receiver knows perfect channel state information and there is a noisy feedback from the receiver, the diversity is limited by the noise in the feedback [6]. Further, if the transmitter learns the channel state through a quantized noiseless link and the receiver does not know the channel state information, the diversity is limited by the noise in the forward channel and the quantization [6]. Even if the feedback link is perfect and has infinite capacity, the training cannot be sent on the feedback channel since the training will indicate about the quality of the feedback channel rather than the forward channel. Hence the best estimate that the transmitter can receive is the noisy estimate from the receiver. On the contrary, in TDD systems, infinite diversity can be achieved if the receiver or the transmitter knows the channel state perfectly. This is shown in Appendix D.

## B. Iterative Protocol with (K-1).5 Rounds

In this Section, we will provide an iterative protocol for a TDD system considering the noise in the training and the feedback channels. This protocol achieves better diversity multiplexing tradeoff than the known results even for the special case of $K = 2$. We will now prove a Lemma that will be used to analyze our proposed iterative protocol.

**Lemma 1.** *Suppose that the transmitter is trained with a power of* $\mathsf{SNR}^p$ *for* $p \geq 1$, *so that the transmitter estimates the channel* $\widehat{H}_1$ *with the eigenvalues of* $\widehat{H}_1 \widehat{H}_1^\dagger$ *being* $\doteq \mathsf{SNR}^{-\widehat{\alpha}_i}$. *Further, the transmit power for the training signal from the transmitter is* $P(\widehat{H}_1) \doteq \frac{\mathsf{SNR}^{1-\epsilon/mn}}{\prod_{i=1}^{mN} \mathsf{SNR}^{-(2i-1+|n-m|)\widehat{\alpha}_i}}$. *Then, the probability that the channel cannot support the rate of* $R$ *on this power controlled channel is asymptotically equivalent to*

$$\Pi(\log\det(I + P(\widehat{H}_1) H H^\dagger) < R) \doteq \mathsf{SNR}^{-mnp - G(r, 1+(mn-1)p)} = \mathsf{SNR}^{-mn(1+pmn)+(m+n-1)r}.$$

*Proof:* This case results in transmitter knowing noisy CSIT, and hence the result is similar in nature to that in [10]. Although the result is similar in nature, we provide the proof in Appendix E for completion. ∎

**Remark 4.** We note from the proof of Lemma 1 that the outage event (event that $\log\det(I + P(\widehat{H}_1) H H^\dagger) < R$) has exponentially small probability given $\alpha_m < p$. So, the outage happens when all the $\alpha_i \geq p$, or in other words when all the eigen-values of $HH^\dagger$ are in the bad state. Further, in this bad state the channel estimation does not work well in the sense that given that $\widehat{\alpha}_i \geq p$, all one can state about $\alpha_i$ is that $\alpha_i \geq p$ with probability one. For example, if $\widehat{\alpha}_i \geq 100p$, all one can reliably say about $\alpha_i$ is that it is $\geq p$. In SISO case, the above property of $\alpha_i$ implies that the channel cannot be resolved below the noise floor since the noise dominates the training signal. The interesting implication in MIMO is that this result of noise dominance holds for all the eigen-values. None of the eigen-value of the channel can be resolved beyond $\alpha_i \geq p$ if $\widehat{\alpha}_i \geq p$. Raising the power exponent $p$ for the feedback reduces the probability of the bad state ($\widehat{\alpha}_m \geq p$) thus increasing diversity.

We now describe the iterative, multi-round TDD protocol. The basic idea of the protocol lies in Lemma 1. If a higher power is allowed on the feedback or training or both, higher diversity can be obtained. However, if higher power is used for all transmissions, then the nodes will violate their power constraint. Hence, we iteratively refine the information about the channel and use higher power for the feedback on-demand. Since the large powers are used rarely, both the nodes stay within their prescribed long-term power constraints.

Let $W_1(r) = 0$, $W_2(r) = mn + G(r, mn)$, $W_k(r) = mn(1 + W_{k-1}(r)) - \epsilon$ for $k > 2$.

1) **Round 1(forward):** The transmitter remains silent.
   **Round 1(reverse):** The receiver sends the training signal using power SNR, $X_{f,1} = \sqrt{\mathsf{SNR}}\beta$, where $\beta$ is a known fixed sequence. The signal received at $\mathsf{T}$ is $Y_{f,1} = H_f X_{f,1} + W$, which is used to form an MMSE estimate $\widehat{H}_1$. Let the eigenvalues of $\widehat{H}_1 \widehat{H}_1^\dagger$ be $\widehat{\lambda}_{1,i} \doteq \mathsf{SNR}^{-\widehat{\alpha}_{1,i}}$. Further, the transmitter will save a quantization index of the channel which is an estimate based on the received power and is labeled $\widehat{q}_{u-1}(Y_{f,i})$ at the end of round $u - 1$. To state the recursion, we assume $\widehat{q}_1 = 1$.

2) **Round** $u \in \{2, K-1\}$**(forward):** The transmitter trains the receiver with $X_u = \sqrt{P_{u-1}(\widehat{H}_{u-1})}\beta$ where the power level $P(\widehat{H}_{u-1})$ is chosen as follows

$$P_{u-1}(\widehat{H}_{u-1}) = \begin{cases} 0, & \text{if } \widehat{q}_{u-1} = 0 \\ \frac{\mathsf{SNR}}{\prod_{i=1}^{mN} \mathsf{SNR}^{-(2i-1+|n-m|)\widehat{\alpha}_{u-1,i}}}, & \text{otherwise.} \end{cases} \quad (22)$$

Note that for power constraint, SNR in the numerator can be replaced by $\mathsf{SNR}^{1-\delta/mn}$ for $\delta$ very small and that will not affect the analysis as was seen in the proof of Lemma 1.

If $u = 2$, the receiver estimates the power controlled channel $G_2 = \sqrt{\widehat{P_1(\widehat{H}_1)}}H$ and decides an index $q_2(G_2)$ as

$$q_2(G_2) = \begin{cases} 0, & \text{if } G_2 \in \mathcal{O}_2^c \\ 1, & \text{otherwise,} \end{cases} \quad (23)$$

where $\mathcal{O}_2 = \{G_2 : \det(I + G_2 G_2^\dagger) < \mathsf{SNR}^{r+\epsilon}\}$.

For $u > 2$, the receiver's actions can be described as follows.
- If $q_{u-1} = 0$, $q_u(q^{u-1}, G_u) = 0$.
- If $q_{u-1} = 1$, a MAP power estimation is done by the receiver. If the power $\dot{<}\mathsf{SNR}^{\epsilon/2mn}$, $q_u = 1$. Otherwise, the receiver estimates the power controlled channel $G_u = \sqrt{\widehat{P(\widehat{H}_{u-1})}}H$, and

$$q_u(q^{u-1}, G_u) = \begin{cases} 0, & \text{if } G_u \in \mathcal{O}_u^c \\ 1, & \text{otherwise,} \end{cases} \quad (24)$$

where $\mathcal{O}_u = \{G_u : \det(I + G_u G_u^\dagger) < \mathsf{SNR}^{r+\epsilon}\}$.

**Round** $u \in \{2, K-1\}$ **(reverse):** The receiver sends the signal

$$X_{f,u} = \begin{cases} 0 \cdot \beta_f, & q_u = 0 \\ \sqrt{\mathsf{SNR}^{1+W_u(r+\epsilon)}}\beta_f, & q_u = 1 \end{cases}, \quad (25)$$

where $\beta_f$ is a known sequence. The signal received at node T is $Y_{f,u} = H_f X_{f,u} + W_f$.

The transmitter estimates the index sent by the receiver using estimate of the received power with a threshold of $\mathsf{SNR}^{\epsilon/mn}$. If the received power $\dot{>}\mathsf{SNR}^{\epsilon/mn}$, then $\widehat{q}_u = 1$ else $\widehat{q}_u = 0$. If $\widehat{q}_u = 1$, the transmitter estimates the channel as $\widehat{H}_{u-1}$. Let the eigenvalues of $\widehat{H}_{u-1}\widehat{H}_{u-1}^\dagger$ be $\widehat{\lambda}_{u-1,i} \doteq \mathsf{SNR}^{-\widehat{\alpha}_{u-1,i}}$.

3) **Round** $K$ **(forward only):** The transmitter trains the receiver and then sends the data both at a power level of

$$P_K(\widehat{H}_1, \cdots, \widehat{H}_{K-1}) = \mathsf{SNR}^{\max(1+\sum_{i=1}^{\min(m,n)}(2i-1+|n-m|)\widehat{\alpha}_{1,i}, \max_{u \in [2, K-1]: \widehat{q}_u = 1} 1+\sum_{i=1}^{\min(m,n)}(2i-1+|n-m|)\widehat{\alpha}_{u,i})} \quad (26)$$

**Theorem 5.** *If the iterative TDD protocol is used for $K \geq 2$, following diversity-multiplexing tradeoff can be achieved*

$$d_{(K-1).5} = \begin{cases} mn\frac{(mn)^K - 1}{mn - 1} - (mn)^{K-2}(m+n-1)r & mn > 1 \\ K - r & mn = 1 \end{cases}.$$

*Proof:* We will show in Appendix F that diversity of $mn(1 + W_{K-1}(r+\epsilon))$ can be obtained, which for $\epsilon \to 0$ converges to the statement of the Theorem. ∎

**Remark 5.** As a special case of the theorem, for $0 < r < \min(m,n)$, diversity of $mn + G(r, mn) = mn(mn+1) - (m+n-1)r$ can be obtained in the TDD system with the 1.5 rounds of training and feedback. In FDD system, the transmitter learns constant number of bits about the channel estimate of the forward channel (channel from the transmitter to the receiver) in the first round. However in the TDD case, the transmitter knows the forward channel estimate (due to reciprocity in the two channels) with larger resolution. Hence, even though the transmitter remains silent in the first round, it receives more information in the first round. From the point of view of the receiver, only the power-controlled channel estimate in the second round is used for decoding the data since it has atleast the same information about the channel (asymptotically) as the first round of training and hence the first training is neglected for decoding. Hence, the TDD protocol performs better than its FDD counterpart. However, the TDD strategy results in the same diversity multiplexing tradeoff performance for a SIMO/MISO system as the FDD strategy when the number of feedback levels $K \to \infty$.

In the iterative TDD protocol, we used a training symbol from the transmitter in each round. However, we could have used one bit of data from the transmitter in each round (since the protocol is receiver initiated, this is like the quantized feedback from the transmitter.). One bit of quantized feedback can be used to achieve a diversity of $G_K(r)$. Although $G_K(r)$ is smaller than what we achieve with this iterative TDD protocol, it takes less time to send one bit of data (one channel use) than to train ($m$ channel uses). Thus when accounting for the resources used in training and feedback, an optimization over the use of a quantized feedback symbol or the training symbol in each round of communication would also be required.

## C. Discussion of TDD Protocols

Like in the iterative protocol for FDD systems, we extend the TDD protocol to an iterative protocol where higher powers are used rarely leading to arbitrarily large gains in diversity with increase in communication rounds. The TDD protocols use reciprocity in addition to the asymmetric properties of the power control. Due to the channel symmetry of the forward and the backward channels, the receiver will be able to find whether the transmitter made an error in understanding the power level. Thus, the channel symmetry can be used to increase the diversity order beyond that is achievable using FDD protocols.

Channel reciprocity enables the receiver to detect if the transmitter has made an error in understanding the feedback signal sent by the receiver. Thus, if the receiver indicated that the channel was Bad in the previous feedback phase but the transmitter understood it as Good, then transmitter's actions will allow the receiver to detect this error with high probability. In this case, the receiver can re-send its feedback signal with higher power.

To understand the specific operation of the protocol, we will use the pictorial view depicted in Figure 5 for a SISO system with 3.5 rounds of TDD protocol. After the forward phase of second round, the receiver knows if the channel is Good (outside the circle in Figure 5(a)) or Bad (inside the circle), and indicates its first feedback as $q_2$. Since the feedback channel is same as the forward channel, the transmitter can make an error in understanding the feedback about only the Bad channel ($q_2 = 1$). That is because when the forward channel is Good, so is the reverse channel. Thus, the feedback in this case is received with exponentially small error probability. On the other hand, when the forward channel is Bad, indicated by $q_2 = 0$, so is the feedback channel. In this case, the transmitter will make errors ($\widehat{q}_2 = 0$ when $q_2 = 1$) if the actual channel was inside the smaller circle of Figure 5(b), while not make errors in the shell between the two circles since the channel is not that bad ($\widehat{q}_2 = 1$ when $q_2 = 1$). In the inner most circle, since $q_2 = 1$, the receiver knows (with exponentially small error probability) that the channel is Bad but the transmitter send no more training assuming it is a Good channel since it thinks that no more channel resolution is required. The receiver can detect the erroneous training power and realize that the transmitter made an error. Detecting an error, the receiver will re-send the same information as $q_3 = 1$ but with higher power to ensure safer delivery. For clarity, we step through each phase of the protocol.

In the first round, the receiver sends a training to the transmitter based on which the transmitter finds the estimated channel. Using this estimated channel, the transmitter trains the receiver in the forward phase of second round. With this training, the receiver finds the regions where this power level will be sufficient to avoid outage. It sends $q_2 = 0$ when the channel is Good (the event that the channel is not in outage based on the estimate of power-controlled channel) while $q_2 = 1$ when it is Bad (not Good). Thus, the channel space is divided in two parts at the receiver. This is depicted in Figure 5(a). The receiver sends this index to the transmitter and due to the asymmetry in the feedback errors, Good channel state is received as Good state ($\widehat{q}_2 = 0$) with probability 1. However, the Bad channel state ($q_2 = 1$) may be mistakenly understood as Good channel state ($\widehat{q}_2 = 0$) by the transmitter. Hence, as shown in Figure 5(b), the transmitter receives the index in error inside the inner circle.

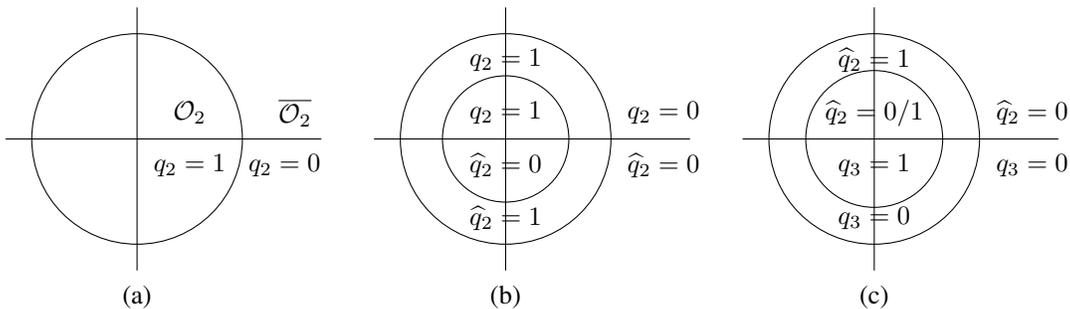

Fig. 5. Channel events (a) in the second round (forward), (b) in the second round (reverse) and (c) in the third round (forward).

In the third round, the transmitter sends a training symbol if it understood that the channel is in Bad state ($\widehat{q}_2 = 1$) which is in the shell between the two circles. Note that no training is sent in the inner circle where the transmitter received incorrect feedback. Since outside the outer circle, the receiver has resolved the channel as being Good, it continues to inform Good channel state ($q_3 = 0$). In the interior of the inner circle where the transmitter was mistaken, the receiver will be able to know that the transmitter made an error (This is because the no training is sent and the power-controlled forward channel will be in outage, i.e. $\log(1 + GG^\dagger) < \mathsf{SNR}^{r+\epsilon}$ where $G$ is the power-controlled channel estimate, with probability 1). This is due to the symmetry which allowed receiver to know that the transmitter has made an error and the channel is Bad. Hence, the receiver informs the transmitter that the channel is Bad ($q_3 = 1$) in this region. However, when the receiver sent Bad channel in the previous round which is correctly decoded as Bad at the transmitter ($q_2 = \widehat{q}_2 = 1$), the receiver will make an estimate and decide if this power level is enough to avoid outage. Based on this, it classifies the region into Good ($q_3 = 0$) or Bad ($q_3 = 1$) as can be seen in Figure 5(c). Outside the outermost circle, the transmitter receives Good ($\widehat{q}_3 = 0$). In the shell between the two circles also, the transmitter receives Good ($\widehat{q}_3 = 0$) due to the asymmetry in the feedback errors. However, the inside circle would be divided in two parts since the transmitter may mistake the Bad channel state ($q_3 = 1$) as Good channel state. The inner most would be received as Good ($\widehat{q}_3 = 0$) while the outer part will be received as Bad ($\widehat{q}_3 = 1$). Thus, we see that the region corresponding to the uncertain part at the transmitter is always a circle centered at the origin and whose radius keeps on shrinking with more rounds of the iterative protocol.

There are two main outage events in this protocol. The first outage event is the channel state being in the inner most circle where the transmitter received Good ($\widehat{q}_3 = 0$) since lower power would be used for transmission than is needed to avoid outage. The second outage event is the area that the receiver classified as Bad ($q_3 = 1$) and the transmitter correctly decoded it as Bad ($\widehat{q}_3 = 1$) and the higher power level used by the transmitter is insufficient to avoid outage. Since both these outage events can be represented as one inner-circle which shrinks with increasing rounds, the resulting diversity order increases in accordance.

Note that both FDD and TDD protocols have the element of using information collected in previous rounds to form the next feedback signal. This memory in encoding channel state information is critical to *adaptive zooming* into the actual channel state. As apparent from the above discussion, the protocol zooms into the poor channel states on-demand, and at the same time tries to keep both the transmitter and receiver aligned in their information about the channel.

## V. Conclusions

In this paper, we derived the diversity tradeoff for a FDD system and a TDD system in which the errors in channel estimation and the feedback channel are fully accounted. This paper finds the diversity multiplexing tradeoffs when the receiver and the transmitter are allowed to exchange multiple rounds of messages. The diversity multiplexing tradeoff increases with each round of message passing between the nodes increases the achievable diversity multiplexing. This paper gives a way of resolving errors with more rounds of message passing in a system where the errors in training and feedback are considered.

The two models, FDD and TDD, considered in this paper are the two extreme cases of the correlations between the forward and the backward channel. As a next step, one can consider what happens if the forward and the feedback channel are correlated, but not identical as in our TDD protocol. When the channels have any constant correlation, the results of FDD system will apply since any constant correlation will make the two channels look independent in high SNR regime. Further analysis to consider the case when the correlation between the forward and the backward channel $\rho$ that satisfies $1 - \rho^2 \doteq \mathsf{SNR}^{-u}$ for any $u > 0$ would be a transition from the FDD approaches to the TDD approaches on increasing $u$, and is still open.

Also, this paper assumes a Rayleigh fading channel model. The authors of [27] consider a general model for fading which includes Rayleigh, Rician, Nakagami and Weibull distributions to find the diversity multiplexing tradeoff for a system with no feedback and perfect channel estimate at the receiver. The extension of the feedback cases to general fading models is still open.

## VI. ACKNOWLEDGEMENTS

The authors wish to thank Gajanana Krishna for useful discussions related to Lemma 1.

## APPENDIX A
### TECHNICAL PRELIMINARIES

*1) Distributions of eigen-value exponents with training:* Suppose that the actual channel is $H$. Let that eigenvalues of $HH^\dagger$ be $(\lambda_1, \cdots, \lambda_{m_N})$, where $m_N = \min(m,n)$. Without loss of generality, we will use $m_N = m$ throughout this paper. For more generality, $m$ can be replaced by $\min(m,n)$ and $n$ by $\max(n,m)$. Further, let $\lambda_i \doteq \mathsf{SNR}^{-\alpha_i}$ and $\boldsymbol{\alpha} = (\alpha_1, \cdots, \alpha_{m_N})$. The distribution of $\alpha_i$'s is given as follows.

**Lemma 2.** *[25] Assume $\alpha_1 \geq \alpha_2 \geq \cdots \alpha_{m_N}$ are the power exponents as described above. In the limit of high* SNR*, the probability density function of the* SNR *exponents, $\boldsymbol{\alpha}$, of the eigenvalues of $HH^\dagger$ is given by*

$$p(\boldsymbol{\alpha}) \doteq \Pi_{i=1}^{m_N} \mathsf{SNR}^{-(2i-1+|n-m|)\alpha_i} \mathbf{1}_{\min(\boldsymbol{\alpha}) \geq 0}. \tag{27}$$

Now suppose that the actual channel $H$ is estimated via a training symbol at a power level of $\mathsf{SNR}^p$. The MMSE channel estimate is denoted by $\widehat{H}$. Let the eigenvalues of $\widehat{H}\widehat{H}^\dagger$ be $(\widehat{\lambda}_1, \cdots, \widehat{\lambda}_{m_N})$, $\widehat{\lambda}_i \doteq \mathsf{SNR}^{-\widehat{\alpha}_i}$ and $\widehat{\boldsymbol{\alpha}} = (\widehat{\alpha}_1, \cdots, \widehat{\alpha}_{m_N})$.

Let $\alpha_1 \geq \alpha_2 \geq \cdots \alpha_{m_N}$ and $\widehat{\alpha}_1 \geq \widehat{\alpha}_2 \geq \cdots \widehat{\alpha}_{m_N}$ (Note that throughout the paper, this ordering will be assumed without loss of generality). Further, we define

$$E_k = \{(\boldsymbol{\alpha}, \widehat{\boldsymbol{\alpha}}) : \min(\alpha_i, \widehat{\alpha}_i) \geq p \; \forall i = 1, \cdots, k, \text{ and } 0 \leq \alpha_i = \widehat{\alpha}_i < p \; \forall \, i > k\} \tag{28}$$

for all $0 \leq k \leq m_N$.

**Lemma 3.** *Let $H$ be the channel and $\widehat{H}$ be the estimated channel using training power of* $\mathsf{SNR}^p$*. In the limit of high* SNR*, the probability density function of the* SNR *exponents of the eigenvalues of $HH^\dagger$ and $\widehat{H}\widehat{H}^\dagger$ is given by*

$$p(\boldsymbol{\alpha}, \widehat{\boldsymbol{\alpha}}) \doteq \sum_{k=0}^{m_N} e_k \mathbf{1}_{E_k} \tag{29}$$

*where $\alpha_1 \geq \alpha_2 \geq \cdots \alpha_{m_N}$, $\widehat{\alpha}_1 \geq \widehat{\alpha}_2 \geq \cdots \widehat{\alpha}_{m_N}$, and*

$$e_k = \mathsf{SNR}^{kp(|n-m|+k)} \Pi_{i=1}^k \mathsf{SNR}^{-(2i-1+|n-m|)\widehat{\alpha}_i} \Pi_{i=1}^{m_N} \mathsf{SNR}^{-(2i-1+|n-m|)\alpha_i}. \tag{30}$$

*Proof:* This lemma was proved in [6] for $p=1$ and is straightforward to generalize for any $p > 0$. ∎

**Corollary 3.** *For the case of $m=n=1$, $p(\alpha, \widehat{\alpha}) \doteq \mathsf{SNR}^{-\alpha} \mathbf{1}_{0 \leq \alpha = \widehat{\alpha} < p} + \mathsf{SNR}^{p-\alpha-\widehat{\alpha}} \mathbf{1}_{\min(\alpha,\widehat{\alpha}) \geq p}$.*

Note from above that if $\alpha_i < p$, $\widehat{\alpha}_i \neq \alpha_i$ with $\doteq 0$ probability. Thus, the eigen-value exponent can be assumed to be correct based on the training if it is $< p$. However if it is $\geq p$, then the value of $\alpha_i$ and $\widehat{\alpha}_i$ have no relation except that $\widehat{\alpha}_i \geq p$. Thus, only limited information can be extracted. This is because the estimation noise becomes dominant.

A novel scheme called power-controlled training was suggested in [4] which takes care of the estimation error in training by estimating the power-controlled channel. If the actual channel is $H$ and the transmit power is $\mathsf{SNR}^{p_j}$. Suppose that the receiver do not know $p_j$. The receiver makes an estimate of $\sqrt{\mathsf{SNR}^{p_j}} H$. With this estimate, the estimation error given by trace $\left(\mathsf{SNR}^{p_j} \tilde{H}\tilde{H}^\dagger\right)$ will be at-most of the order of constant. This is because the channel estimation error in $H$, $\tilde{H}$ have complex normal entries with 0 mean and variance $\mathsf{SNR}^{p_j}$. Thus, let eigen-values of $\mathsf{SNR}^{p_j} \tilde{H}\tilde{H}^\dagger$ be $\tilde{\lambda}_i \doteq \mathsf{SNR}^{-\tilde{\alpha}_i}$ where $\tilde{\alpha}_i$ have $\doteq 0$ probability if $\alpha_i < 0$ as given before. Then, $\text{trace}(\mathsf{SNR}^{p_j} \tilde{H}\tilde{H}^\dagger) \doteq \mathsf{SNR}^{-\tilde{\alpha}_m} \dot{\leq} \mathsf{SNR}^0$, Thus, the estimation error is at the noise floor. For more details, the reader is referred to [4,6]. This scheme will be used in all the protocols in this paper.

*2) Power Controlled Feedback:* We will use the concept of power-controlled feedback [6] throughout the paper. For illustration, we will focus on a case of distinguishing between two feedback levels. Consider feedback levels $a$ and $b$ are transmitted from the receiver using power levels of $\mathsf{SNR}^{p_1}$ and $\mathsf{SNR}^{p_2}$ respectively. Without loss of generality, assume that $p_1 < p_2$. The transmitter will observe the received power and find if the receiver sent $a$ or $b$. As in [6], if the receiver transmitted a symbol at $\mathsf{SNR}^{p_i}$, the received power is $\doteq \mathsf{SNR}^{(p_i - \alpha_m)^+}$ where $\alpha_m$ is the smallest eigen-value exponent of the channel as was defined in the last subsection. We will use the threshold detection at the transmitter. In [6], threshold of $\mathsf{SNR}^{p_1+\delta}$ was proven to be optimal for arbitrarily small $\delta > 0$. To see this, we consider the error events at the transmitter:

1) $\Pi(\text{Transmitter received } a | \text{Receiver transmitted } b) \doteq \Pi(p_2 - \alpha_m < p_1 + \delta) \doteq \mathsf{SNR}^{-mn(p_2-p_1-\delta)}$.

2) $\Pi(\text{Transmitter received } b | \text{Receiver transmitted } a) \doteq \Pi(p_1 - \alpha_m > p_1 + \delta) \doteq 0$.

Since the second event happens with probability $\doteq 0$, $\delta$ is chosen arbitrarily close to 0 to get the optimal detection at the transmitter.

## APPENDIX B
## PROOF OF THEOREM 2

We will first verify that the average power constraint is satisfied.

1) **Round 1**: For the feedback, $c_M \leq 1$ and thus $(c_M - G_u(r+\epsilon))^+ \leq 1$. Further,
$\Pi\left(q_1(\widehat{H}_1) = K-1\right) = \Pi\left(\det(I + \widehat{H}_1\widehat{H}_1^\dagger \mathsf{SNR}^{1+p_{K-2}}) < \mathsf{SNR}^{r+\epsilon}\right) \doteq \mathsf{SNR}^{-G(r+\epsilon, 1+p_{K-2})}$. Thus, the power constraint in feedback is satisfied.

2) **Round 2**: To observe the satisfiability of the power constraint, we will find the probabilities of $q_1(\widehat{H}_1)$ and $\widehat{q}_1$.
We note that the probability of $q_1(\widehat{H}_1)$ is bounded as:

$$\Pi\left(q_1(\widehat{H}_1) = u\right) \leq \mathsf{SNR}^{-\min(G_u(r+\epsilon), c_M)} = \mathsf{SNR}^{-p_u} \text{ if } u < K-1 \tag{31}$$

$$\Pi\left(q_1(\widehat{H}_1) = K-1\right) \leq \mathsf{SNR}^{-G(r+\epsilon, 1+p_{K-2})} \tag{32}$$

The first term is because if $G_{u-1}(r+\epsilon) < c_M$ then $\Pi\left(q_1(\widehat{H}_1) = u\right) \doteq \mathsf{SNR}^{G_u(r+\epsilon)}$, but if $G_{u-1}(r+\epsilon) > c_M$ then $\Pi\left(q_1(\widehat{H}_1) = u\right) \doteq 0$.

Since the received power cannot have a larger SNR exponent than the transmitted power asymptotically,

$$\Pi(\widehat{q}_1 = K-1) \doteq \Pi\left(q_1(\widehat{H}_1) = K-1\right) \leq \mathsf{SNR}^{-G(r+\epsilon, 1+p_{K-2})}. \tag{33}$$

For $0 \leq u < K-1$,

$$\Pi(\widehat{q}_1 = u) = \Pi\left(\widehat{q}_1 = u, q_1(\widehat{H}_1) \geq u\right) + \Pi\left(\widehat{q}_1 = u, q_1(\widehat{H}_1) < u\right) \tag{34}$$

$$\dot{\leq} \mathsf{SNR}^{-p_u} + \Pi\left(\widehat{q}_1 = u, q_1(\widehat{H}_1) < u\right) \tag{35}$$

The last step follows since $\Pi\left(q_1(\widehat{H}_1) = u_2 \geq u\right) \dot{\leq} \mathsf{SNR}^{-p_u}$. Thus,

$$\Pi(\widehat{q}_1 = u) \dot{\leq} \mathsf{SNR}^{-p_u} + \Pi\left(\widehat{q}_1 = u, q_1(\widehat{H}_1) < u\right) \tag{36}$$

$$\dot{\leq} \mathsf{SNR}^{-p_u} + \sum_{i=0}^{u-1} \Pi\left(\widehat{q}_1 = u | q_1(\widehat{H}_1) = i\right) \Pi\left(q_1(\widehat{H}_1) = i\right) \tag{37}$$

$$\dot{\leq} \mathsf{SNR}^{-p_u} + \sum_{i=0}^{u-1} \Pi\left(\widehat{q}_1 = u | q_1(\widehat{H}_1) = i\right) \mathsf{SNR}^{-p_i} \tag{38}$$

$$\dot{\leq} \mathsf{SNR}^{-p_u} + \sum_{i=0}^{u-1} \mathsf{SNR}^{-mn\left((c_M - G_i(r+\epsilon))^+ - (c_M - G_u(r+\epsilon))^+ - \delta/mn\right)} \mathsf{SNR}^{-p_i} \tag{39}$$

$$\dot{\leq} \mathsf{SNR}^{-p_u} + \sum_{i=0}^{u-1} \mathsf{SNR}^{-mn(\min(c_M, G_u(r+\epsilon)) - G_i(r+\epsilon))^+ + \delta - \min(c_M, G_i(r+\epsilon))} \tag{40}$$

$$\dot{\leq} \mathsf{SNR}^{-p_u} + \sum_{i=0}^{u-1} \mathsf{SNR}^{-(p_u - \delta)^+} \tag{41}$$

$$\dot{\leq} \mathsf{SNR}^{-(p_u - \delta)^+} \tag{42}$$

Hence, the power constraint is satisfied. We will now use $\delta \approx 0$ to compute the outage probability of different events.

1) $\det(I + HH^\dagger \mathsf{SNR}) \geq \mathsf{SNR}^r$. In this case, any power level is sufficient, and hence there is no outage.

2) For $1 \leq u < K-1$, $\det\left(I + HH^\dagger \mathsf{SNR}^{1+\min(c_M, G_{u-1}(r+\epsilon))}\right) < \mathsf{SNR}^r$ and $\det\left(I + HH^\dagger \mathsf{SNR}^{1+\min(c_M, G_u(r+\epsilon))}\right) \geq \mathsf{SNR}^r$. We will first show that in this case $q_1(\widehat{H}_1) \geq u$. Let the eigenvalues of $HH^\dagger$ and $\widehat{H}_1\widehat{H}_1^\dagger$ be $\mathsf{SNR}^{-\alpha_i}$ and $\mathsf{SNR}^{-\widehat{\alpha}_i}$ respectively. Note that

$$\Pi\left(\det(I + HH^\dagger\mathsf{SNR}^{1+\min(c_M, G_{u-1}(r+\epsilon))}) < \mathsf{SNR}^r,\right.$$
$$\left.\det(I + \widehat{H}_1\widehat{H}_1^\dagger\mathsf{SNR}^{1+\min(c_M, G_{u-1}(r+\epsilon))}) \geq \mathsf{SNR}^{r+\epsilon}\right) \tag{43}$$

$$\doteq \Pi\left(\sum_{i=1}^m (1 + p_{u-1} - \alpha_i)^+ < r, \sum_{i=1}^m (1 + p_{u-1} - \widehat{\alpha}_i)^+ \geq r + \epsilon\right). \tag{44}$$

If $r \leq m-1$, $p_{u-1} = 0$ and thus $\sum_{i=1}^m (1 + p_{u-1} - \alpha_i)^+ = \sum_{i=1}^m (1 + p_{u-1} - \widehat{\alpha}_i)^+$ making the above probability $\doteq 0$. For $r > m-1$, first consider that atleast $k \geq 1$ $\alpha_i$'s are $\geq 1$ (or, $\alpha_1, \cdots, \alpha_k \geq 1$). In this case,

$$\sum_{i=1}^m (1 + p_{u-1} - \widehat{\alpha}_i)^+ \leq kp_{u-1} + (m-k)(1 + p_{u-1}) \tag{45}$$
$$\leq kc_M + (m-k)(1 + c_M) \tag{46}$$
$$\leq mc_M + m - k \tag{47}$$
$$\leq r - (m-1) + m - 1 \tag{48}$$
$$\leq r. \tag{49}$$

Thus, $\sum_{i=1}^m (1 + p_{u-1} - \widehat{\alpha}_i)^+ \geq r + \epsilon$ can never happen. Thus, only possibility is that all $\alpha_i < 1$ in which case $\sum_{i=1}^m (1 + p_{u-1} - \alpha_i)^+ = \sum_{i=1}^m (1 + p_{u-1} - \widehat{\alpha}_i)^+$. Since $q_1(\widehat{H}_1) \geq u$, $\widehat{q}_1 \geq u$ and thus there is no outage in this case.

3) $\det\left(I + HH^\dagger\mathsf{SNR}^{1+\min(c_M, G_{K-2}(r+\epsilon))}\right) < \mathsf{SNR}^r$ and $\det\left(I + HH^\dagger\mathsf{SNR}^{1+\min(G_{K-1}(r+\epsilon), G(r, 1+c_M))}\right) \geq \mathsf{SNR}^r$. As before, $q_1(\widehat{H}_1) \neq K - 1$ with probability $\doteq 0$. However, $\widehat{q}_1$ can now take a lower value. Thus, the outage happens whenever any less power is received at the transmitter. This happens when the power of $\mathsf{SNR}^{1+G(r+\epsilon, 1+p_{K-2})}$ is received below $\mathsf{SNR}^{c_u+\delta/mn}$ which happens with probability $\mathsf{SNR}^{-mn(1+G(r+\epsilon, 1+p_{K-2})-c_M)+\delta}$.

4) $\det\left(I + HH^\dagger\mathsf{SNR}^{1+\min(G_{K-1}(r+\epsilon), G(r, 1+c_M))}\right) < \mathsf{SNR}^r$. This happens with probability $\mathsf{SNR}^{-G(r, 1+\min(G_{K-1}(r+\epsilon), G(r, 1+c_M)))}$.

Thus, the overall outage probability is bounded by

$$\Pi(\mathcal{O}) \dotlesssim \mathsf{SNR}^{-mn(1+G(r+\epsilon, 1+p_{K-2})-c_M)+\delta} + \mathsf{SNR}^{-G(r, 1+\min(G_{K-1}(r+\epsilon), G(r, 1+c_M)))} \tag{50}$$
$$\doteq \mathsf{SNR}^{-mn(1+G(r+\epsilon, 1+p_{K-2})-c_M)+\delta} + \mathsf{SNR}^{-\min(G(r, 1+G_{K-1}(r+\epsilon)), G(r, 1+G(r, 1+c_M)))} \tag{51}$$

For $r \leq m-1$, $p_{K-2} = c_M = 0$ and thus, $mn(1+G(r+\epsilon, 1+p_{K-2})-c_M)+\delta = mn(1+G(r+\epsilon, 1))+\delta \geq G(r, 1+G(r+\epsilon, 1))$. Thus, choosing $\epsilon$ and $\delta$ close to 0, diversity of $\min(G_K(r), G(r, 1+G(r, 1+c_M)))$ can be obtained.

For $r > m-1$, Let $d = G(r+\epsilon, 1+p_{K-2})$. We will show that $mn(1+d-c_M) \geq G(r, 1+d)$. With this, the diversity will reduce to $\min(G_K(r), G(r, 1+G(r, 1+c_M)))$.

To see $mn(1+d-c_M) \geq G(r, 1+d)$, note that for $m = 1$, left and right sides are both $n(1+d-r)$ and thus the inequality is satisfied. For $m > 1$,

$$G(r, 1+d) \leq G(1, 1+d) \tag{52}$$
$$= mn(1+d) - (m+n-1) \tag{53}$$
$$\leq mn(1+d) - n \tag{54}$$
$$\leq mn(1+d) - mnc_M \tag{55}$$

## APPENDIX C
## PROOF OF THEOREM 3

We will first verify that the power levels used in the protocol satisfy the average power constraints.

1) **Round 1**: Transmitter trains receiver with power $\mathsf{SNR}$, and hence the average power constraint is satisfied. The receiver uses a maximal power level of $\mathsf{SNR}$ and hence the average power constraint is satisfied.
2) **Round $2 \leq u+1 \leq K-1$**:

$$\Pi(\widehat{q}_1 = 1) = \Pi(\widehat{q}_1 = 1, q_1(\widehat{H}_1) = 0) + \Pi(J_{\mathsf{T}1} = 1, q_1(\widehat{H}_1) = 1)$$
$$\leq \Pi(\widehat{q}_1 = 1 | q_1(\widehat{H}_1) = 0) + \Pi(q_1(\widehat{H}_1) = 1)$$
$$\dotlesssim \mathsf{SNR}^{-mn+\epsilon} + \mathsf{SNR}^{-G_1(r+\epsilon)}$$
$$\doteq \mathsf{SNR}^{-G_1(r+\epsilon)}, \tag{56}$$

where we used $\Pi(\widehat{q}_1 = 1 | q_1(\widehat{H}_1) = 0) \doteq \mathsf{SNR}^{-mn+\epsilon}$ since the threshold is at $\mathsf{snr}^{\epsilon/mn}$ and by [6]. Further, $\Pi(q_1(\widehat{H}_1) = 1) = \det(I + \widehat{H}_1 \widehat{H}_1^\dagger \mathsf{SNR}) < \mathsf{SNR}^{r+\epsilon} \doteq \mathsf{SNR}^{-G_1(r+\epsilon)}$. The last step follows since $G_1(r+\epsilon) \leq G_1(\epsilon) = mn - \epsilon(m+n-1) \leq mn - \epsilon$.

Hence, the power constraint is satisfied at the transmitter for $u = 1$.

$$
\begin{aligned}
\Pi(q_{u+1} = 0) &= \Pi(q_{u+1} = 0, q_u = 0) + \Pi(q_{u+1} = 0, q_u = 1) \\
&\doteq \Pi(q_{u+1} = 0, q_u = 0, \widehat{q}_u = 0) + \Pi(q_{u+1} = 0, q_u = 0, \widehat{q}_u = 1) \\
&\quad + \Pi(q_{u+1} = 0, \widehat{q}_u = 1, q_u = 1) \\
&\doteq \Pi(q_{u+1} = 0, q_u = 0, \widehat{q}_u = 1) + \Pi(q_{u+1} = 0, \widehat{q}_u = 1, q_u = 1) \\
&\dot\leq \Pi(\widehat{q}_u = 1) \\
&\dot\leq \mathsf{SNR}^{-G_u(r+\epsilon)}
\end{aligned} \tag{57}
$$

where we used three things:

a) $\Pi(q_{u+1} = 0, q_u = 1, \widehat{q}_u = 0) \doteq 0$. If $u = 1$, $\Pi(q_u = 1, \widehat{q}_u = 0) \doteq 0$. For this consider two cases. If $q_i = 0$ for some $i < u$, then as $\widehat{q}_u = 0$, $q_{u+1} = 1$ since the power estimate $\geq \mathsf{SNR}^\epsilon$ happens with probability $\doteq 0$. Further, if $q_i = 1$ for all $i \leq u$, $\widehat{q}_u = 0$ happens with probability $\doteq 0$ since any transmitted 1 from the receiver gets received as 0 with exponentially small probability.

b) $\Pi(q_{u+1} = 0, q_u = 0, \widehat{q}_u = 0) \doteq 0$. This is because when transmitter sends at 0 power, the receiver receives at power $\geq \mathsf{SNR}^\epsilon$ with exponentially small probability.

c) $\Pi(\widehat{q}_u = 1) \dot\leq \mathsf{SNR}^{-G_u(r+\epsilon)}$ which has been earlier shown to be true for $u = 1$, and we will show in general by induction.

Hence, the average power constraint at the receiver is satisfied for all $u$. To check the induction argument for $\widehat{q}_u$ and transmit power constraint,

$$
\Pi(\widehat{q}_{u+1} = 1) = \Pi(\widehat{q}_{u+1} = 1, q_{u+1} = 0) + \Pi(\widehat{q}_{u+1} = 1, q_{u+1} = 1). \tag{58}
$$

For the first term $\Pi(\widehat{q}_{u+1} = 1, q_{u+1} = 0)$. This happens when $\mathrm{trace}(HH^\dagger)\mathsf{SNR}^{1+G_u(r+\epsilon)} + 1 \dot\leq \mathsf{SNR}^{\epsilon/mn}$ which happens with probability $\mathsf{SNR}^{-mn(1+G_u(r+\epsilon))+\epsilon}$.

The second term can occur only when all the $q_1$ to $q_{u+1} = 1$, and all $\widehat{q}_1$ to $\widehat{q}_{u+1} = 1$. The reason for all $q_1$ to $q_{u+1} = 1$ is because if any of $q_i = 0$ all $q_{i+1} = q_{u+1} = 0$ since the forward channel was good and cannot be bad in the next rounds. This happens when $\det(I + \widehat{H}_{u+1} \widehat{H}_{u+1}^\dagger \mathsf{SNR}^{1+G_u(r+\epsilon)}) < \mathsf{SNR}^{r+\epsilon}$, which happens with probability $\mathsf{SNR}^{-G_{u+1}(r+\epsilon)}$.

Note that

$$
\begin{aligned}
G_{u+1}(r+\epsilon) &= G(r+\epsilon, 1 + G_u(r+\epsilon)) \\
&\leq G(\epsilon, 1 + G_u(r+\epsilon)) \\
&= mn(1 + G_u(r+\epsilon)) - (m+n-1)\epsilon \\
&\leq mn(1 + G_u(r+\epsilon)) - \epsilon.
\end{aligned} \tag{59}
$$

Hence,

$$
\Pi(\widehat{q}_{u+1} = 1) \dot\leq \mathsf{SNR}^{-G_{u+1}(r+\epsilon)}. \tag{60}
$$

Hence, the induction steps hold, and the power constraint is satisfied.

3) **Round $K$ (forward only)**:

$$
\begin{aligned}
\Pi(\widehat{q}_1 = 1, \widehat{q}_2 = 1) &= \Pi(\widehat{q}_1 = 1, \widehat{q}_2 = 1, q_2 = 0) + \Pi(\widehat{q}_1 = 1, \widehat{q}_2 = 1, q_1 = 0, q_2 = 1) \\
&\quad + \Pi(\widehat{q}_1 = 1, \widehat{q}_2 = 1, q_1 = 1, q_2 = 1) \\
&\dot\leq \Pi(\widehat{q}_1 = 1, \widehat{q}_2 = 1, q_2 = 0, q_1 = 0) \\
&\quad + \Pi(\widehat{q}_1 = 1, \widehat{q}_2 = 1, q_2 = 0, q_1 = 1) \\
&\quad + \Pi(q_1 = 0, \widehat{q}_1 = 1, q_2 = 1) + \Pi(q_2 = 1 | q_1 = 1, \widehat{q}_1 = 1).
\end{aligned} \tag{61}
$$

The first term $\widehat{q}_1 = 1, \widehat{q}_2 = 1, q_2 = 0, q_1 = 0$ happens asymptotically with probability that there is error in the reverse channel both times. This happens with probability $\mathsf{SNR}^{-mn(1+G_1(r+\epsilon))+\epsilon}$.

The second term $\widehat{q}_1 = 1, \widehat{q}_2 = 1, q_2 = 0, q_1 = 1$ happens when $q_1 = 1$ and when there is error in the feedback channel in the second round. This happens with probability less than $\mathsf{SNR}^{-mn(1+G_1(r+\epsilon))+\epsilon}$.

The third term $q_1 = 0, \widehat{q}_1 = 1, q_2 = 1$ happens with exponentially small probability since the probability $\det(I + H_1 H_1^\dagger \mathsf{SNR}) \geq \mathsf{SNR}^{r+\epsilon}$ and $\mathrm{trace}(H_2 H_2^\dagger \mathsf{SNR}^{1+G_1(r+\epsilon)}) + 1 < \mathsf{SNR}^\epsilon$ is $\doteq 0$. Note that $i, j^{\mathrm{th}}$ term of $H_1$ and $H_2$ are correlated with correlation $\rho_{i,j}$ s.t. $1 - \rho_{i,j}^2 \doteq \mathsf{SNR}^{-1}$. Let the eigenvalues of $H_1 H_1^\dagger$ and $H_2 H_2^\dagger$ be $\lambda_1, \cdots, \lambda_{\min(m,n)}$ and $\mu_1, \cdots, \mu_{\min(m,n)}$ respectively. Let $\lambda_i = \mathsf{SNR}^{\alpha_i}$ and $\mu_i = \mathsf{SNR}^{-\widehat{\alpha_i}}$.

$$\begin{aligned}
&\Pi\left(\det(I + H_1 H_1^\dagger \mathsf{SNR}) \geq \mathsf{SNR}^{r+\epsilon}, \operatorname{trace}(H_2 H_2^\dagger \mathsf{SNR}^{1+G_1(r+\epsilon)}) + 1 < \mathsf{SNR}^\epsilon\right) \\
&\doteq \Pi\left(\sum(1-\alpha_i)^+ \geq r+\epsilon, (1+G_1(r+\epsilon) - \min_i \widehat{\alpha}_i)^+ < \epsilon\right) \\
&\doteq \Pi\left(\sum(1-\widehat{\alpha}_i)^+ \geq r+\epsilon, (1+G_1(r+\epsilon) - \min_i \widehat{\alpha}_i)^+ < \epsilon\right) \\
&\doteq 0
\end{aligned} \quad (62)$$

The fourth term in right hand side of (61),

$$\Pi(q_2 = 1 | q_1 = 1, \widehat{q}_1 = 1) \doteq \mathsf{SNR}^{-G_2(r+\epsilon)}. \quad (63)$$

Hence, (61) reduces to

$$\begin{aligned}
\Pi(\widehat{q}_1 = 1, \widehat{q}_2 = 1) &\doteq \mathsf{SNR}^{-mn(1+G_1(r+\epsilon))+\epsilon} + \mathsf{SNR}^{-G_2(r+\epsilon)} \\
&\doteq \mathsf{SNR}^{-G_2(r+\epsilon)}.
\end{aligned} \quad (64)$$

Similarly, we see that for $u \geq 2$, if $\widehat{q}_1 = \cdots = \widehat{q}_u = 1$. Then, if any of $q_u = i$ for some $i \leq u$, then $q_u = 0$. Also, the feedback channel transforms $q_u = 0$ to $\widehat{q}_u = 1$, which happens with probability $\mathsf{SNR}^{-mn(1+G_{u-1}(r+\epsilon))+\epsilon}$. However, if $q_i = 1$ for all $i \leq u$, this happens when $\log \det(I + H_u H_u^\dagger \mathsf{SNR}^{1+G_{u-1}(r+\epsilon)}) \geq \mathsf{SNR}^{r+\epsilon}$ which happens with probability $\mathsf{SNR}^{-G_u(r+\epsilon)}$. Hence, the event $\widehat{q}_1 = \cdots = \widehat{q}_u = 1$ happens with probability $\doteq \mathsf{SNR}^{-G_u(r+\epsilon)}$. Hence, the power constraint is satisfied.

We will now show that the outage probability is $\doteq \mathsf{SNR}^{-G(r,1+G_{K-1}(r+\epsilon))}$. For this, we see the region of $H$ when there will be outage.

1) $\det(I + HH^\dagger \mathsf{SNR}) \geq \mathsf{SNR}^r$. In this case, power of $\mathsf{SNR}$ is sufficient, while we will be using a higher power level. Hence, there can be no outage.
2) $\det(I + HH^\dagger \mathsf{SNR}) < \mathsf{SNR}^r$ and $\det(I + HH^\dagger \mathsf{SNR}^{1+G_1(r+\epsilon)}) \geq \mathsf{SNR}^r$. In this case, $q_1 = 1$ with probability 1. The reason is that $\Pi\left(\sum_{i=1}^{\min(m,n)}(1-\alpha_i)^+ \leq r, \sum_{i=1}^{\min(m,n)}(1-\widehat{\alpha}_i)^+ \geq r+\epsilon\right) \doteq 0$ (where $\alpha_i$ and $\widehat{\alpha}_i$ are the negative exponents of the eigenvalues of $HH^\dagger$ and $\widehat{H}_1 \widehat{H}_1^\dagger$). Since $q_1 = 1$, $\widehat{q}_1 = 1$ with probability 1. Since $\widehat{q}_1 = 1$, power of atleast $\mathsf{SNR}^{2-r-\epsilon}$ is used and hence no outage in this range.
3) For $u \in [2, K-1]$, $\det(I + HH^\dagger \mathsf{SNR}^{1+G_{u-1}(r+\epsilon)}) < \mathsf{SNR}^r$ and $\det(I + HH^\dagger \mathsf{SNR}^{1+G_u(r+\epsilon)}) \geq \mathsf{SNR}^r$. As before, $q_1 = \widehat{q}_1 = 1$. Now, since training is at $\mathsf{SNR}^{1+G_1(r+\epsilon)}$, $\Pi\left(\sum_{i=1}^{\min(m,n)}(1+G_1(r+\epsilon)-\alpha_i)^+ \leq r,\right.$
$\left.\sum_{i=1}^{\min(m,n)}(1+G_1(r+\epsilon)-\widehat{\alpha}_i)^+ \geq r+\epsilon\right) \doteq 0$ (where $\alpha_i$ and $\widehat{\alpha}_i$ are the negative exponents of the eigenvalues of $HH^\dagger$ and $\widehat{H}_2 \widehat{H}_2^\dagger$). Hence, $q_2 = 1$ which implies $\widehat{q}_2 = 1$ with probability 1. Similarly extending all till $q_u = \widehat{q}_u = 1$ with probability 1. Hence there is no outage in this range

Hence, the outage happens only when $\det(I + HH^\dagger \mathsf{SNR}^{1+G_u(r+\epsilon)}) < \mathsf{SNR}^r$ which happens with probability $\mathsf{SNR}^{-G(r,1+G_{K-1}(r+\epsilon))}$. For $\epsilon$ very small, diversity of $G_K(r)$ can be achieved.

## APPENDIX D
## TDD: PERFECT CSIR OR PERFECT CSIT

In this Appendix, we consider two cases. The first case assumes perfect channel state information at the transmitter, where only receiver is unaware of channel conditions. Second, we consider the case of perfect information at the receiver, with no channel information at the transmitter. In both the above cases, infinite diversity order can be achieved with 1.5 round protocols which is unlike the case of FDD systems[6].

For the case of perfect channel knowledge at the transmitter and receiver, it is well known [1] for the case of $\min(m,n) > 1$, zero outage probability is possible at a finite SNR value and thus the diversity order is unbounded. For the case of single antenna system, i.e $\min(m,n) = 1$, channel inversion leads to an exponential decay in probability and hence also has an infinite diversity order. We show in the next two results that infinite diversity order can be achieved even if only one of the nodes, transmitter or receiver has perfect channel information.

We begin the discussion with the case when the channel information is perfect at the transmitter and receiver relies on estimated channel knowledge.

**Lemma 4** (No CSIR, Perfect CSIT). *For all $0 < r < \min(m,n)$, an infinite diversity order is achievable with a 0.5 round protocol.*

*Proof:* The protocol proceeds as follows.

**Round 1 (forward only):** The transmitter trains the receiver with a power level

$$P(H) \doteq \frac{\mathsf{SNR}^{1-\epsilon/mn}}{\Pi_{i=1}^{m_N}\mathsf{SNR}^{-(2i-1+|n-m|)\alpha_i}},$$

where $\lambda_i \doteq \mathsf{SNR}^{-\alpha_i}$ are the eigen-values of the channel realization $H$ at the transmitter, and sends data at the same power. It is straight-forward to conclude that the power constraint will be satisfied, and the outage probability $\doteq 0$ for $r < \min(m,n) - \epsilon$. (By choosing $\epsilon$ small enough, the above holds for $r < \min(m,n)$.) Note that the power control from the transmitter is only asymptotically satisfied. ∎

Since the transmitter knows the channel perfectly, it can "invert" the channel exactly for the case of $\min(m,n) > 1$ and in this case, the receiver sees a channel whose SNR does not fluctuate and is always above the threshold needed to avoid the outage. Similarly, for the case of no information at the transmitter but full information at the receiver, we can achieve an error probability decay which is faster than any polynomial decay in SNR.

**Lemma 5** (Perfect CSIR, No CSIT). *For $0 < r < \min(m,n)$, an infinite diversity order is achievable with a 1.5 rounds protocol.*

*Proof:* The protocol proceeds as follows.

1) **Round 1 (forward):** The transmitter remains silent.
   **Round 1 (reverse):** Since the receiver knows the channel $H$, it finds the index of the power level from $i \in [1, K]$ that would be sufficient to avoid outage as in [8]. Let $\lambda_i \doteq \mathsf{SNR}^{-\alpha_i}$ are the eigen-values of $H$ with $\alpha_1 \geq \alpha_2 \geq \cdots \geq \alpha_m$. Further, the receiver communicates this power level to the transmitter using a power level of $\mathsf{SNR}^{\alpha_m + \frac{i+1}{2K}}$ to communicate index $i$. The transmitter estimates the received power and hence gets the index $i$ with error probability $\doteq 0$.
2) **Round 2 (forward only):** Since the transmitter now knows the power level to use, it sends data at the required power level.

First note that the probability with which the power level of $\mathsf{SNR}^{\alpha_m + \frac{i+1}{2K}}$ is used is $\dot{\leq}\mathsf{SNR}^{-\alpha_m}$ and hence the power constraint is satisfied. The only step to prove is to show that the probability with which the transmitter will make a mistake to decode the index transmitted by the receiver is $\doteq 0$. Suppose that the transmitter uses a threshold for level $i$ at $\mathsf{SNR}^{\frac{i+1}{2K} + \frac{1}{4K}}$. Since the transmitted power is $\mathsf{SNR}^{\alpha_m + \frac{i+1}{2K}}$, the received power is $\doteq \mathsf{SNR}^{\alpha_m + \frac{i+1}{2K}} \mathsf{SNR}^{-\alpha_m} = \mathsf{SNR}^{\frac{i+1}{2K}}$. Here, we exploited the reciprocity in the channels since $\alpha_m$ for the forward and the backward channels are the same. Since there is nothing random in the above exponent of SNR in this received power (asymptotically), using the above threshold, the index be mistaken with probability $\doteq 0$.

The above strategy attains the diversity of $G_K(r)$ for any finite $K$. Since $G_K(r)$ monotonically increases with $K$ for $0 < r < \min(m,n)$, for any given $x \geq 0$ diversity $\geq x$ can be achieved by choosing $K$ large enough. Thus, the above protocol yields unbounded diversity. ∎

# APPENDIX E
# PROOF OF LEMMA 1

The probability that the channel cannot support rate $R$ is given by

$$\Pi\left(\log\det(I + P(\widehat{H}_1)HH^\dagger) < R\right)$$

$$\doteq \Pi\left(\sum_{i=1}^{\min(m,n)}(1 - \frac{\epsilon}{mn} + \sum_{i=1}^{\min(m,n)}(2i-1+|n-m|)\widehat{\alpha}_i - \alpha_i)^+ < r\right)$$

$$\doteq \int_{\mathcal{A}} p(\boldsymbol{\alpha},\widehat{\boldsymbol{\alpha}})d\boldsymbol{\alpha}d\widehat{\boldsymbol{\alpha}}$$

$$\doteq \sum_{k=0}^{\min(m,n)} \int_{\mathcal{A}\cap E_k} \mathsf{SNR}^{k(|n-m|+k)}\Pi_{i=1}^k \mathsf{SNR}^{-(2i-1+|n-m|)\widehat{\alpha}_i}\Pi_{i=1}^{m_N}\mathsf{SNR}^{-(2i-1+|n-m|)\alpha_i}d\boldsymbol{\alpha}d\widehat{\boldsymbol{\alpha}}, \qquad (65)$$

where $\mathcal{A} = \{\boldsymbol{\alpha},\widehat{\boldsymbol{\alpha}} : \sum_{i=1}^{\min(m,n)}\left(1 - \frac{\epsilon}{mn} + \sum_{i=1}^{\min(m,n)}(2i-1+|n-m|)\widehat{\alpha}_i - \alpha_i\right)^+ < r\}$

We first note that $\mathcal{A} \cap E_k = \Phi$ for $k < \min(m,n)$. Let $m \leq n$, other case is similar. For any $E_k$ where $k < \min(m,n)$, $0 \leq \alpha_i = \widehat{\alpha}_i < p \forall i > k$ and $\min(\alpha_i, \widehat{\alpha}_i) \geq p \forall i = 1, \cdots, k$. Thus,

$$\sum_{i=1}^{\min(m,n)} \left(1 - \frac{\epsilon}{mn} + \sum_{i=1}^{\min(m,n)} (2i - 1 + |n - m|)\widehat{\alpha}_i - \alpha_i\right)^+$$

$$= \sum_{i=1}^{k} \left(1 - \frac{\epsilon}{mn} + \sum_{i=1}^{\min(m,n)} (2i - 1 + |n - m|)\widehat{\alpha}_i - \alpha_i\right)^+$$

$$+ \sum_{i=k+1}^{m} \left(1 - \frac{\epsilon}{mn} + \sum_{i=1}^{k}(2i - 1 + |n - m|)\widehat{\alpha}_i + \sum_{i=k+1}^{m} (2i - 1 + |n - m|)\alpha_i - \alpha_i\right)^+$$

$$\geq \sum_{i=k+1}^{m} \left(1 - \frac{\epsilon}{mn} + \sum_{i=1}^{k}(2i - 1 + |n - m|)\widehat{\alpha}_i + \sum_{i=k+1}^{m} (2i - 1 + |n - m|)\alpha_i - \alpha_i\right)^+$$

$$\geq \sum_{i=k+1}^{m} \left(1 - \frac{\epsilon}{mn} + \sum_{i=1}^{k}(2i - 1 + |n - m|)\widehat{\alpha}_i\right)^+$$

$$\geq \sum_{i=k+1}^{m} \left(1 - \frac{\epsilon}{mn} + \sum_{i=1}^{k}(2i - 1 + |n - m|)\right)^+$$

$$= (m - k)\left(1 - \frac{\epsilon}{mn} + k(k + |n - m|)\right)$$

$$\geq (m - k)\left(1 + k^2 - \frac{\epsilon}{mn}\right) \tag{66}$$

For $k = 0$, above is $\geq m(1 - \epsilon/mn)$ and for any $r < m - \epsilon$, the above cannot be less than $r$. Similar for $k = 1, m > 1$, above cannot be $< r$. For $k > 1$ we can lower bound $k^2 - k - \frac{\epsilon}{mn}$ by 0 (by choosing $\epsilon < 1$). Thus,

$$\sum_{i=1}^{\min(m,n)} \left(1 - \frac{\epsilon}{mn} + \sum_{i=1}^{\min(m,n)} (2i - 1 + |n - m|)\widehat{\alpha}_i - \alpha_i\right)^+$$

$$\geq (m - k)\left(1 + k^2 - \frac{\epsilon}{mn}\right)$$

$$\geq (m - k)(1 + k)$$

$$= m - k + mk - k^2$$

$$= m + (m - 1 - k)k$$

$$\geq m \tag{67}$$

Hence,

$$\Pi\left(\log\det(I + P(\widehat{H}_1)HH^\dagger) < R\right)$$

$$\doteq \int_{\mathcal{A} \cap E_m} \mathsf{SNR}^{mp(n-m+m)} \Pi_{i=1}^{m} \mathsf{SNR}^{-(2i-1+n-m)\widehat{\alpha}_i} \Pi_{i=1}^{m} \mathsf{SNR}^{-(2i-1+n-m)\alpha_i} d\boldsymbol{\alpha} d\widehat{\boldsymbol{\alpha}}$$

$$\doteq \int_{\mathcal{A} \cap E_m} \mathsf{SNR}^{mnp} \Pi_{i=1}^{m} \mathsf{SNR}^{-(2i-1+n-m)\widehat{\alpha}_i} \Pi_{i=1}^{m} \mathsf{SNR}^{-(2i-1+n-m)\alpha_i} d\boldsymbol{\alpha} d\widehat{\boldsymbol{\alpha}} \tag{68}$$

Hence, the negative of the SNR exponent of the above probability is

$$\min_{(\boldsymbol{\alpha},\widehat{\boldsymbol{\alpha}}) \in \mathcal{A} \cap E_m} \sum_{i=1}^{m}(2i - 1 + n - m)(\widehat{\alpha}_i + \alpha_i) - mnp. \tag{69}$$

The optimal $\widehat{\alpha}_i = p$. Now, let $\alpha_i = p + \beta_i$. The above negative SNR exponent is given by

$$mnp + \min_{(\beta_1,\cdots,\beta_m) \in \mathcal{B}} \sum_{i=1}^{m}(2i - 1 + n - m)\beta_i \tag{70}$$

where $\mathcal{B} = \{(\beta_1, \cdots, \beta_m) : \beta_i \geq 0, \sum_{i=1}^{m}(1 + (mn-1)p - \frac{\epsilon}{mn} - \beta_i)^+ < r$. By the definition of function $G$, the above reduces to

$$mnp + G(r, 1 + (mn - 1)p - \frac{\epsilon}{mn}). \tag{71}$$

For $\epsilon \to 0$, the probability that the channel cannot support rate $R$ is thus $\mathsf{SNR}^{-mnp-G(r,1+(mn-1)p-\frac{\epsilon}{mn})}$.

# APPENDIX F
## PROOF OF THEOREM 5

To verify that the power constraint is satisfied, observe the following.

1) **Round 1:** Receiver trains the transmitter with power SNR, and hence power constraint is satisfied.
2) **Round 2:** The transmitter trains the receiver with a power level $P_1\left(\widehat{H}_1\right) \doteq \frac{\mathsf{SNR}}{\Pi_{i=1}^{mN} \mathsf{SNR}^{-(2i-1+|n-m|)\widehat{\alpha}_{1,i}}}$, and the average power constraint is satisfied as was shown in single round.

$$\Pi(q_2 = 1) = \Pi\left(\det\left(I + G_2 G_2^\dagger\right) \geq \mathsf{SNR}^{r+\epsilon}\right) \tag{72}$$
$$\dot{\leq} \mathsf{SNR}^{-W_2(r+\epsilon)}, \tag{73}$$

where the second steps follows from Lemma 1 with $p = 1$.

3) **Round** $u = 3 : K - 1$:

$$\Pi(\widehat{q}_2 = 1) = \Pi(\widehat{q}_2 = 1, q_2 = 1) + \Pi(\widehat{q}_2 = 1, q_2 = 0) \tag{74}$$
$$\dot{\leq} \Pi(q_2 = 1) + \Pi(\widehat{q}_2 = 1 | q_2 = 0) \tag{75}$$
$$\dot{\leq} \mathsf{SNR}^{-W_2(r+\epsilon)} + e^{-\mathsf{SNR}^\epsilon} \tag{76}$$
$$\dot{=} \mathsf{SNR}^{-W_2(r+\epsilon)} \tag{77}$$

Further, the training power from the transmitter satisfies the power constraint by the same arguments as in the single round,

$$\Pi(q_u = 1) = \Pi(q_u = 1, \widehat{q}_{u-1} = 1, q_{u-1} = 1) + \Pi(q_u = 1, \widehat{q}_{u-1} = 0, q_{u-1} = 1). \tag{78}$$

Let us first consider the first term. As $q_{u-1} = 1$ is received as $\widehat{q}_{u-1} = 1$, $\alpha_m \leq 1 + W_{u-1}(r+\epsilon) - \frac{\epsilon}{mn}$. Since $\alpha_m$ is less than the SNR exponent of the power at which the training symbol was sent, $G_u \in \mathcal{O}_u$ with probability $\dot{=} 0$ by Lemma 1. Also, the power estimation cannot fail since $\alpha_m = \widehat{\alpha}_m$. Thus, the first term is $\dot{=} 0$.

The second term is upper bounded by $\Pi(\widehat{q}_{u-1} = 0 | q_{u-1} = 1)$ which happens with probability $\mathsf{SNR}^{-mn(1+W_{u-1}(r+\epsilon))+\epsilon} \dot{=} \mathsf{SNR}^{-W_u(r+\epsilon)}$. Hence,

$$\Pi(q_u = 1) = \Pi(q_u = 1, \widehat{q}_{u-1} = 1, q_{u-1} = 1) + \Pi(q_u = 1, \widehat{q}_{u-1} = 0, q_{u-1} = 1) \tag{79}$$
$$\dot{\leq} \mathsf{SNR}^{-W_u(r+\epsilon)}. \tag{80}$$

The power constraint at the transmitter is satisfied since the average power is bounded by SNR.

4) **Round** $K$ **(forward only)**: The transmitter trains the receiver and then sends the data both at a power level of

$$\left\{\mathsf{SNR}^{\max\left(1+\sum_{i=1}^{\min(m,n)}(2i-1+|n-m|)\widehat{\alpha}_{1,i}, \max_{u \in [2, K-1] : \widehat{q}_u = 1}\left(1+\sum_{i=1}^{\min(m,n)}(2i-1+|n-m|)\widehat{\alpha}_{u,i}\right)\right)}\right\} \tag{81}$$

Note that the power constraint is satisfied since for any maximizing term in the power, the probability that that particular $\widehat{q} = 1$ would balance the exponent.

We will now show that the outage probability is $\dot{=} mn(1 + W_{k-1}(r+\epsilon))$. For $K = 2$, it follows directly by Lemma 1 for p=1. We now assume that $K > 2$. For this, we see the region of $H$ when there will be outage.

1) $\det\left(I + HH^\dagger P_1\left(\widehat{H}_1\right)\right) \geq \mathsf{SNR}^r$. In this case, power of $P\left(\widehat{H}_1\right)$ is sufficient, while we will be using a higher power level. Hence, there can be no outage.
2) $\det\left(I + HH^\dagger P_1\left(\widehat{H}_1\right)\right) < \mathsf{SNR}^r$ and $\alpha_m < 1 + W_2(r+\epsilon) - \frac{\epsilon}{mn}$. In this case, $q_2 = 1$ with probability 1. The reason is that $\Pi\left(\sum_{i=1}^{\min(m,n)}(p-\alpha_i)^+ \leq r, \sum_{i=1}^{\min(m,n)}\left(p-\widehat{\alpha}_i\right)^+ \geq r+\epsilon\right) \dot{=} 0$ (where $\alpha_i$ and $\widehat{\alpha}_i$ are the negative exponents of the eigenvalues of $HH^\dagger$ and $H_2 H_2^\dagger$, where $H_2$ is the non-power controlled channel estimate at the receiver and $p$ is the SNR exponent in $P\left(\widehat{H}_1\right)$). Further, $\widehat{q}_2 = 1$ because $\alpha_m < 1 + W_2(r+\epsilon) - \frac{\epsilon}{mn}$ makes the received power $\dot{>} \mathsf{SNR}^{\epsilon/mn}$. As $\widehat{q}_2 = 1$, using power $P\left(\widehat{H}_2\right)$ does not produce any outage when $\alpha_m < 1 + W_2(r+\epsilon) - \frac{\epsilon}{mn}$.
3) For $u \in [3, K-1]$, $\det\left(I + HH^\dagger P_1\left(\widehat{H}_1\right)\right) < \mathsf{SNR}^r$, $1 + W_{u-1}(r+\epsilon) - \frac{\epsilon}{mn} \leq \alpha_m < 1 + W_{u-1}(r+\epsilon) - \frac{\epsilon}{mn}$. As before, $q_2 = 1$. Now, since $\alpha_m \geq 1 + W_2(r+\epsilon) - \frac{\epsilon}{mn}$, the transmitter will receive this index as 0 which the receiver would know. Hence, $q_3 = 1$. This continues till $q_u = 1$. However, $\widehat{q}_u = 1$ since $\alpha_m < 1 + W_{u-1}(r+\epsilon) - \frac{\epsilon}{mn}$. Moreover, sending at power $P_u\left(\widehat{H}_u\right)$ when $\alpha_m < 1 + W_{u-1}(r+\epsilon) - \frac{\epsilon}{mn}$ does not produce any outage.

Hence, the outage happens only when $\det\left(I + HH^\dagger P_1\left(\widehat{H}_1\right)\right) < \mathsf{SNR}^r$ and $\alpha_m \geq 1 + W_{u-1}(r+\epsilon) - \frac{\epsilon}{mn}$, whose probability is less than the probability of $\alpha_m \geq 1 + W_{K-1}(r+\epsilon) - \frac{\epsilon}{mn}$ which happens with probability $\mathsf{SNR}^{-W_K(r+\epsilon)}$. For $\epsilon$ very small, the diversity multiplexing tradeoff as in the statement of the Theorem can be achieved.

## REFERENCES

[1] E. Biglieri, G. Caire and G. Taricco, "Limiting performance for block-fading channels with multiple antennas," *IEEE Transactions on Information Theory*, vol. 47, no. 4, pp. 1273-1289, May 2001.
[2] A. Khoshnevis and A. Sabharwal, "On the asymptotic performance of multiple antenna channels with quantized feedback," *IEEE Transactions on Wireless Communications*, 10 (7), pp. 3869 - 3877, October 2008.
[3] V. Sharma, K. Premkumar and R. N. Swamy, "Exponential diversity achieving spatio-temporal power allocation scheme for fading channels," *IEEE Transactions on Information Theory*, vol. 54, no. 1, Jan. 2008.
[4] C. Steger and A. Sabharwal, "Single-input two-way SIMO channel: Diversity-multiplexing tradeoff with two-way training," *IEEE Transactions on Wireless Communications*, vol. 7, no. 12, pp. 4877-4885, Dec 2008.
[5] G. G. Krishna, S. Bhashyam and A. Sabharwal, "Decentralized power control with two-way training for multiple access," in *Proc. IEEE International Symposium on Information Theory*, July 2008, Toronto.
[6] V. Aggarwal and A. Sabharwal, "Power-controlled feedback and training for two-way MIMO channels," *Submitted to IEEE Transactions on Information Theory,* Jan. 2009.
[7] S. Ekbatani, F. Etemadi and H. Jafarkhani, "Outage behavior of slow fading channels with power control using noisy quantized CSIT," *arXiv:0804.0790v1*, Apr. 2008.
[8] T. T. Kim and M. Skoglund, "Diversity-multiplexing tradeoff in MIMO channels with partial CSIT," *IEEE Transactions on Information Theory*, vol. 53, Issue 8, pp. 2743-2759, Aug. 2007.
[9] V. Aggarwal and A. Sabharwal, "Performance of multiple access channels with asymmetric feedback," *IEEE Journal on Selected Areas in Communication*, vol. 26, no. 8, pp. 1516-1525, Oct. 2008.
[10] T. T. Kim and G. Caire, "Diversity gains of power control with noisy CSIT in MIMO channels," *IEEE Transactions on Information Theory*, vol. 55, pp. 1618-1626, Apr. 2009..
[11] V. Aggarwal and A. Sabharwal, "Diversity order gain with noisy feedback in multiple access channels," in *Proc. IEEE International Symposium on Information Theory*, July 2008, Toronto.
[12] A. Khoshnevis and A. Sabharwal, "Achievable diversity and multiplexing in multiple antenna systems with quantized power control," *in Proc. IEEE Intl. Conference on Communications,* May 2005, Seoul, Korea.
[13] H. El Gamal, G. Caire, M. O. Damen, "The MIMO ARQ channel: Diversity-multiplexing-delay tradeoff," *IEEE Transactions on Information Theory*, vol. 52, pp. 3601-3621, Aug. 2006.
[14] A. J. Goldsmith and P. P. Varaiya, "Capacity of fading channels with channel side information," *IEEE Transactions on Information Theory,* vol. 43(6), pp. 1986-1992, Nov. 1997.
[15] G. Caire and S. Shamai, "On the capacity of some channels with channel state information," *IEEE Transactions on Information Theory*, vol. 45(6), pp. 2007-2019, Sep. 1999.
[16] S. Bhashyam, A. Sabharwal and B. Aazhang, " Feedback gain in multiple antenna systems," *IEEE Transactions on Comm.*, vol. 50, Issue 5, pp. 785-798, May 2002.
[17] V. Aggarwal, G. G. Krishna, S. Bhashyam and A. Sabharwal, "Two models for noisy feedback in MIMO channels," in *Proc. Asilomar Conference on Signals, Systems and Computers,* Oct. 2008, Pacific Grove, CA.
[18] G. G. Krishna, "Feedback with resource accounting in MIMO systems," *B. Tech. Project Report,* Indian Institute of Technology Madras, May 2008.
[19] A. Narula, M. J. Lopez, M. D. Trott and G. W. Wornell, "Efficient use of side information in multiple-antenna data transmission over fading channels," *IEEE JSAC*, vol. 16, pp. 1423-1436, Oct. 1998.
[20] K. K. Mukkavilli, A. Sabharwal, E. Erkip and B. Aazhang, "On beamforming with finite rate feedback in multiple-antenna systems," *IEEE Transactions on Information Theory*, vol. 49, pp. 2562-2579, Oct. 2003.
[21] C. W. Tan and A. R. Calderbank, "Multiuser detection of Alamouti signals," *IEEE Transactions on Communications,* Vol. 57, No. 7, pp. 2080-2089, Jul. 2009.
[22] J. Schalkwijk and T. Kailath, "A coding scheme for additive noise channels with feedback-I: No bandwidth constraint," *IEEE Transactions on Information Theory,* vol. IT-12, No. 2, pp. 172-182, Apr. 1966.
[23] A. Kramer, "Improving communication reliability by use of an intermittent feedback channel," *IEEE Transactions on Information Theory,* vol. 15, no. 1, pp. 52-60, Jan. 1969.
[24] S. Lavenberg, "Feedback communication using orthogonal signals," *IEEE Transactions on Information Theory,* vol. 15, no. 4, pp. 478-483, Jul. 1969.
[25] L. Zheng and D. N. C. Tse, "Diversity and multiplexing: a fundamental tradeoff in multiple-antenna channels," *IEEE Transactions on Information Theory*, vol. 49, Issue 5, pp. 1073-1096, May 2003.
[26] L. Zheng and D. Tse, "Communicating on the Grassmann manifold: A geometric approach to the non-coherent multiple antenna channel," *IEEE Transactions on Information Theory*, vol. 48(2), pp. 359-383, Feb. 2002.
[27] L. Zhao, W. Mo, Y. Ma and Z. Wang, " Diversity and multiplexing tradeoff in general fading channels," *IEEE Transactions on Information Theory,* vol. 53(4), pp. 1549-1557, Apr. 2007.